\documentclass[a4paper,11pt]{article}
\pdfoutput=1
\usepackage{jcappub}
\usepackage{journal_abbrev}
\usepackage{xparse,verbatim,color}
\usepackage{xspace,ulem,etoolbox,cprotect}
\xspaceaddexceptions{]\}}
	\usepackage{bm}
	\usepackage{hyperref}

\usepackage{graphicx}
\usepackage{adjustbox}
\usepackage{siunitx}
\usepackage{enumitem}
\usepackage{lmodern}

\def\ba#1\ea{\begin{align}#1\end{align}}
\def\bea{\begin{eqnarray}}
\def\eea{\end{eqnarray}}
\def\be{\begin{equation}}
\def\ee{\end{equation}}
\def\d{\delta}

\def\({\left(}
\def\){\right)}
\def\[{\left[}
\def\]{\right]}
\def\<{\left\langle}
\def\>{\right\rangle}
\def\lapl{\nabla^2}
\def\vn{\boldsymbol{\nabla}}
\DeclareMathOperator{\tr}{tr}

\newcommand{\CC}{C\nolinebreak\hspace{-.05em}\raisebox{.4ex}{\tiny\bf +}\nolinebreak\hspace{-.10em}\raisebox{.4ex}{\tiny\bf +}}

\newcommand{\perm}[1]{ \expandafter\ifstrempty\expandafter{#1} {\mbox{perm.}} {\mbox{$#1$ perm.}} }

\newcommand{\vs}{\nonumber\\}
\def\d{{\delta}}
\def\eps{{\varepsilon}}

\renewcommand{\v}[1]{\bm{#1}}

\def\vx{\v{x}}

\def\vk{\v{k}}
\def\vq{\v{q}}

\def\O{\mathcal{O}}

\def\Del{\mathcal{D}}

\def\cvs{\v{t}}
\def\hM{M}

\def\Mpch{\,h^{-1}\text{Mpc}}
\def\iMpch{\,h\,\text{Mpc}^{-1}}

\def\Om{\Omega_m}
\def\cH{\mathcal{H}}
\def\Lbox{L_\text{box}}
\def\L{\Lambda}
\def\Lin{\Lambda_\text{in}}

\def\xfl{\vx_{\rm fl}}

\newcommand{\refeq}[1]{Eq.~(\ref{eq:#1})}

\newcommand{\reffig}[1]{Fig.~\ref{fig:#1}}

\newcommand{\reftab}[1]{Tab.~\ref{tab:#1}}
\newcommand{\refsec}[1]{Sec.~\ref{sec:#1}}

\newcommand{\refapp}[1]{Appendix~\ref{app:#1}}

\def\emph#1{\textit{#1}}

\title{An $\bm{n}$-th order Lagrangian Forward Model for Large-Scale Structure}

\author[a]{Fabian Schmidt}

\emailAdd{fabians@mpa-garching.mpg.de}

\affiliation[a]{Max--Planck--Institut f\"ur Astrophysik, Karl--Schwarzschild--Stra\ss e 1, 85748 Garching, Germany}


\arxivnumber{2012.09837}

\abstract{A forward model of matter and biased tracers at arbitrary order in Lagrangian perturbation theory (LPT) is presented. The forward model contains the complete LPT displacement field at any given order in perturbations, as well as all relevant bias operators at that order and leading order in derivatives. The construction is done for any expansion history and does not rely on the Einstein-de Sitter approximation. A large subset of higher-derivative bias operators is also included. As validation test, we compare the $n$LPT-predicted matter density field and that from N-body simulations using the same initial conditions. For simulations using a cutoff in the initial conditions, we find subpercent agreement up to scales of $k\sim 0.2 \Mpch$. We also find subpercent agreement with full simulations without cutoff, both for the power spectrum and nonlinear $\sigma_8$-inference, when allowing for the effective sound speed. The application to biased tracers (halos) has already been presented in a recent paper \cite{paper_realspace}.}

\begin{document}

\maketitle

\flushbottom

\section{Introduction}
\label{sec:intro}

A \emph{forward model} of large-scale structure (LSS) evolves a given set of
initial conditions---in the form of a linear matter density field used
to initialize the growing mode of structure formation---into the observed
structure today. The latter can consist of the total matter density field,
as probed in projected form via gravitational lensing, or the density
field of biased tracers, as observable using galaxy surveys. Such
forward models can be used for a variety of purposes: to construct mock
observations (see \cite{2019MNRAS.482.1786L} and references therein);
to reduce sample variance in the comparison to full simulation
results \cite{2012JCAP...04..013T,baldauf/LPT,Abidi:2018eyd,Lazeyras:2017hxw,2020arXiv200901200S}; 
to provide an alternative means to predict summary statistics
of the observables such as power spectra and three-point functions (bispectra)
and their covariance \cite{gridSPT,2019MNRAS.482.1786L,2020arXiv200705504T};
and for use in full Bayesian inference approaches, where such forward models
provide a prediction for the mode of the field-level likelihood for given
initial conditions \cite{2013MNRAS.432..894J, 2015MNRAS.446.4250A, 2017JCAP...12..009S, paperII, 2019arXiv190906396L}. 

One well-known forward model of LSS consists in N-body simulations,
which fully nonlinearly solve for the evolution of cold collisionless matter
under gravity. Apart from the fact that full simulations are very costly,
however, perturbative forward models
(see \cite{LSSreview} for a review of perturbative approaches to LSS)
are well motivated by the fact that they allow us to rigorously incorporate
non-gravitational small-scale physics, including baryonic feedback, via
free bias coefficients (see \cite{biasreview} for a review). This
allows them to be applied to the case of biased tracers, whose formation
cannot be simulated realistically in general.

In perturbative forward models, one can take either the Eulerian
\cite{gridSPT} or Lagrangian route \cite{baldauf/LPT,schmittfull/etal:2018,schmittfull/etal:2020}.
However, none of the existing published forward models include the
general bias expansion beyond second order. The forward model presented here aims to add precisely
that. In addition, it also computes the perturbation theory solutions as well
as general bias expansion for
a general expansion history (see also \cite{fujita/vlah,2020JCAP...10..039D}),
rather than the Einstein-de Sitter approximation
assumed in the published perturbative forward models.

The forward model presented here is based on Lagrangian perturbation theory (LPT), which has been studied extensively for matter \cite{1992MNRAS.254..729B,1994MNRAS.267..811B,1995A&A...296..575B,1995MNRAS.276..115C,2012JCAP...06..021R,2015JCAP...09..014V,baldauf/LPT}, but also for biased tracers \cite{matsubara:2008,2020JCAP...07..062C}. There are several strong reasons for using a Lagrangian formulation for LSS forward models:
\begin{enumerate}
\item The LPT prediction shows a much better correlation with the fully nonlinear evolved density field than Eulerian perturbation theory (e.g., \cite{2012JCAP...04..013T,2014JCAP...06..008T}).
\item The general bias expansion described in \cite{MSZ,biasreview} can be implemented in a straightforward way based on $n$-th order LPT. In the Eulerian case, this expansion is much more complicated due to the convective time derivatives involved.
\item Performing the bias expansion first before displacing to the Eulerian position allows for the simple incorporation of redshift-space distortions and lightcone effects (not discussed here in detail, but see \cite{Cabass:2020jqo,schmittfull/etal:2020} and \refsec{conc}).
\end{enumerate}

The general bias expansion, which at any given spacetime location incorporates
the density and tidal field along the past trajectory, is an expansion
in orders of perturbations and derivatives. As mentioned above, the
Lagrangian forward model generates all bias terms up to any given order in LPT,
but at leading order in derivatives. A corresponding expansion up to
any order in derivatives is technically challenging due to memory requirements
and the difficulty in removing degeneracies between different operators.
However, we describe here how to systematically construct a non-degenerate
subset of higher-derivative operators.

The main goal of this paper is to present the forward model and validate
it by comparing the resulting matter density field to that obtained in
full N-body simulations. We also employ N-body simulations that start from
initial conditions with a sharp-$k$ filter, setting all initial perturbations
with $|\vk|>\Lambda$ to zero. This allows for precise comparisons in the
regime where the entire density field remains under perturbative control.

The application to biased tracers has already been presented in \cite{paper_realspace}, based on \cite{cabass/schmidt:2020}. Specifically, that paper
studied the inference of the amplitude of
primordial fluctuations from halos in N-body simulations using the
field-level likelihood based on the effective field theory
\cite{paperI,paperII,paperIIb,cabass/schmidt:2019}.
In that case,
the complete bias expansion including higher-derivative terms described
here was employed. For sufficiently small values of the cutoff $\L$,
convergence to an unbiased estimate was demonstrated. However,
many more tests on biased tracers, including comparing correlation functions,
could be done.

The outline of the paper is as follows: \refsec{LPT} reviews the LPT
construction. We then turn to the general bias expansion in \refsec{bias}
as well as its extension to higher derivatives in \refsec{hderiv}.
The numerical implementation is described in \refsec{code}.
We then present results in \refsec{sim} and conclude in \refsec{conc}.
The appendices give more details on the LPT solution.

\section{Lagrangian perturbation theory}
\label{sec:LPT}

Lagrangian perturbation theory starts with the geodesic equation for
nonrelativistic matter with initially vanishing peculiar
velocity and density perturbation.
We write the spatial part of the solution of this equation of motion as
\be
\xfl(\vq,\tau) = \vq + \v{s}(\vq,\tau),
\label{eq:sdef}
\ee
where $\v{s}(\vq,\tau)$ is the Lagrangian displacement, with
$\lim_{\tau\to 0} \v{s}(\vq,\tau) = 0$, and $\vq$ is the Lagrangian coordinate $\vq$.
Then, the equation for the displacement $\v{s}$ is given by
\be
\v{s}''(\vq,\tau) + \cH \v{s}'(\vq,\tau) = -\vn_x \Phi(\vx,\tau)\Big|_{\vx = \vq+\v{s}(\vq,\tau)},
\label{eq:seom}
\ee
where primes denote derivatives with respect to conformal time $\tau$.
The initial conditions are set at very early times $\tau\to 0^+$ by
setting $\v{s}, \v{s}'$ to a purely longitudinal, growing-mode linear displacement field and its time derivative,
as we will describe below. 
The system is closed via the Poisson equation
\be
\nabla_x^2\Phi(\vx,\tau) = \frac32 \Om(\tau) \cH^2(\tau) \d(\vx,\tau).
\label{eq:Poisson}
\ee
Specifically, we work with a metric in conformal-Newtonian gauge, and with
the matter density defined in synchronous-comoving gauge, so that a fixed
$\tau$ slice corresponds to fixed proper time for comoving observers.
With these definitions, \refeq{Poisson} remains valid on all scales (see Sec.~2.9
of \cite{biasreview} and references therein). Relativistic corrections
then only need to be applied when connecting quantities defined on
proper-time slices to measurements made by a distant observer through
light signals (see \cite{jeong/schmidt:2015} for a review).

Notice further that any scale-independent modification of the linear Poisson equation,
due to a modification of gravity for example, can be incorporated in the following by absorbing the modification into a redefined $\Om(\tau)$.

Two further relations are needed. First, from \refeq{sdef} we have
\be
\partial_{q,i} = \left[\delta_i{}^j + H_i{}^j \right] \partial_{x,j} ,
\quad\mbox{where}\quad
H_{ij}(\vq,\tau) \equiv \partial_{q,i} s_j(\vq,\tau)
\label{eq:Mdef}
\ee
is the Lagrangian distortion tensor (frequently also denoted as $\Psi_{j,i}$).
Second, by choosing the Lagrangian
coordinates such that equal infinitesimal volume intervals $d^3\vq$
contain equal amounts of matter, the matter density contrast $\d$ is
given by
\be
1 + \d(\xfl(\vq,\tau)) = \left|\frac{\partial\vx}{\partial\vq}\right|^{-1}_{\vx=\xfl(\vq,\tau)}
= | \v{1} + \v{H}(\vq,\tau) |^{-1},
\label{eq:dLPT}
\ee
where we denote 3-vectors and 3-tensors in boldface. 
Using \refeq{Mdef} and \refeq{dLPT}, \refeq{seom} can be transformed into
an equation for $\v{H}$ which only refers to Lagrangian coordinates $\vq$:
\be
\tr\left[(\v{1}+\v{H})^{-1} (\v{H}'' + \cH \v{H}')\right] =
-\frac32\Om \cH^2 \left[|\v{1}+\v{H}|^{-1}-1\right].
\label{eq:Meom}
\ee
In the following, we will deal mostly with Lagrangian coordinates, and
correspondingly drop the subscript $q$ on derivatives; derivatives
with respect to Eulerian coordinates $x$ will be indicated explicitly.

LPT now works by treating the components of the distortion tensor as small
parameters \cite{1992MNRAS.254..729B}; this is related to the expansion in powers of $\d$ performed
in Eulerian PT by \refeq{dLPT}.
Specifically, we write
\be
\v{s}(\vq,\tau) = \sum_{n=1}^{\infty} \v{s}^{(n)}(\vq,\tau),
\ee
and similarly for $\v{H}$, where the solution for $\v{s}^{(n)}$ involves
$n$ powers of the linear-order displacement $\v{s}^{(1)}$. 
\refeq{Meom} can then be solved iteratively,
where suitable manipulations show that the source term
only involves quadratic and cubic couplings. This in turn allows for
convenient recursion relations \cite{rampf:2012,zheligovsky/frisch,matsubara:2015}.

First, we decompose the displacement (at any order) into longitudinal and transverse (or curl)
components:
\be
\v{s} = \frac{\vn}{\lapl} \sigma - \frac1{\lapl} \vn\times \cvs,
\label{eq:sdecomp}
\ee
where $\sigma \equiv \vn\cdot\v{s}$ denotes the divergence of the displacement,
which only contributes to the symmetric part of $\v{H}$, 
while the curl $\cvs = \vn\times\v{s}$ also sources 
the antisymmetric part of $\v{H}$. 
Here, we have only allowed for solutions consistent with homogeneity,
  or equivalently periodic boundary conditions, allowing us to
  remove the constant solution and the solution $\propto \vx$ in $\v{s}$ (see \cite{buchert/ehlers:1997}). We will also assume that the spatial average of $\v{s}$ vanishes, likewise in keeping with periodic boundary conditions.

Second, we convert the time coordinate from conformal time $\tau$ to
$\lambda\equiv \ln D$, the logarithm of the linear growth factor which is given
by the growing solution to the linear ODE
\be
D'' + \cH D' - \frac32 \Om \cH^2 D = 0.
\ee
We now follow the treatment of Matsubara \cite{matsubara:2015}.
Combining Eqs.~(26), (28), (31) and (59) there, we then obtain
the following set of coupled ODE describing the evolution of the $n$-th order longitudinal and transverse contributions to $\v{H}$:
\ba
 \Del_{3/2}(\lambda) \sigma^{(n)}(\vq,\lambda) =&
\sum_{m_1+m_2=n} \bigg\{ \tr\left[ \v{H}^{(m_1)}(\vq,\lambda)
  \Del_{3/4}(\lambda) \v{H}^{(m_2)}(\vq,\lambda) \right] \vs
& \hspace*{2cm}- \tr\left[ \v{H}^{(m_1)}(\vq,\lambda) \right]
\Del_{3/4}(\lambda) \tr\left[\v{H}^{(m_2)}(\vq,\lambda) \right]\bigg\} \vs
&
-\frac12 \sum_{m_1+m_2+m_3=n} \eps_{ijk}\eps_{lmn} H^{(m_1)}_{il}(\vq,\lambda) H^{(m_2)}_{jm}(\vq,\lambda)
\Del_{1/2}(\lambda) H^{(m_3)}_{kn}(\vq,\lambda)
\vs
\Del_0(\lambda) (\cvs^{(n)})^i =& \sum_{m_1+m_2=n} \eps^{ijk} \left( \v{H}^{(m_1)} \Del_0 \v{H}^{(m_2)\,\top}\right)_{jk} \;,
\label{eq:eomy}
\ea
where
\ba
\gamma(\lambda) \equiv\:& \frac{\Om(\lambda)}{f^2(\lambda)} - 1; \quad
\lambda\equiv \ln D 
\vs
\Del_c \equiv\:&
\frac{\partial^2}{\partial \lambda^2} + \frac12 [1 + 3\gamma(\lambda)] \frac{\partial}{\partial \lambda}
- c [1+\gamma(\lambda)] \quad\mbox{for any } c\in \mathbb{R}.
\ea
Notice that contributions to the transverse source term with $m_1=m_2$ vanish. For this reason, the transverse component begins at third order. 

In a Euclidean matter-dominated (Einstein-de Sitter, EdS) universe, $\gamma=0$,
and the ODE can be solved analytically for the fastest growing mode. This leads
to the simple recurrence relations given in \refapp{EdS} \cite{zheligovsky/frisch,MSZ}. In particular, both $\sigma$ and $\cvs$ factorize, i.e.
\be
\sigma^{(n)}(\vq,\lambda) = e^{n \lambda} \sigma^{(n)}(\vq,\lambda=0)
\ee
and similarly for the transverse component.

This factorization is essential both for an efficient evaluation of the LPT displacement, and for a closed bias expansion. Fortunately, the factorization is still possible even for a general expansion history \cite{1997GReGr..29..733E}. For this, we write, at fixed order $n$ in perturbations
\ba
\sigma^{(n)}(\vq,\lambda) &= \sum_p [w^L_{n,p} e^{n\lambda} + \alpha^L_{n,p}(\lambda)] \sigma^{(n,p)}(\vq,\lambda=0) \vs
\cvs^{(n)}(\vq,\lambda) &= \sum_p [w^T_{n,p} e^{n\lambda} + \alpha^T_{n,p}(\lambda)] \cvs^{(n,p)}(\vq,\lambda=0),
\label{eq:sepsigmat}
\ea
where $p$ runs over the different shapes relevant at order $n$. Naively, one could
expect one independent shape for each partition of $n$, however the actual number
is significantly smaller due to the source term which only involves quadratic and cubic couplings, which are moreover symmetric in the $m_i$; \reftab{shapes} schematically lists
the different shapes relevant at the first few orders. 
The weights $w^{L,T}_{n,p}$ are set by the EdS solution (\refapp{EdS}), so that in EdS we simply have $\alpha^{L,T}_{n,p}=0$, and all shapes at a given order
have the same time dependence and can be combined into one linear combination.

In the code implementation, the contributions $\sigma^{(n,p)}, \alpha^L_{n,p}$
and $\cvs^{(n,p)}, \alpha^T_{n,p}$ are determined iteratively, starting from the
linear order $n=1$. At a given order $n$ we determine the independent shapes
generated by the source terms in \refeq{eomy}, and integrate the ODE from deep
in matter domination, starting with initial conditions $\alpha^{L,T}_{n,p} = 0 = \alpha^{L,T}_{n,p}{}'$. We only include independent shapes by restricting to $m_2 \geq m_1$ in the quadratic source term, and $m_3 \geq m_2 \geq m_1$ in the cubic source term. 
Note that each of these shapes in general corresponds to two or
three different time dependencies in the source term coming from the permutations among the $m_i$, e.g. from $H^{m_1} \Del H^{m_2}$ and $H^{m_2} \Del H^{m_1}$ if $m_2\neq m_1$. Since they multiply the same $\vq$-dependent shape however, we combine them into a single $\alpha^{L,T}_{n,p}$.

This treatment is valid for any expansion history that has an
extended period of matter domination; no assumption of an expansion history close
to EdS is made at later times.
The number of independent contributions to $\v{H}$ increases rapidly toward
larger $n$; for example, at $n=5$ we have 9 independent longitudinal contributions (and 6 transverse ones), while these numbers grow to 23 (and 14) at $n=6$.
For the fiducial $\Lambda$CDM cosmology,
$\alpha^{L,T}_{n,p}$ grows to at most $\sim 0.002$ at $z=0$.

In the results for this paper, we also neglect the contribution of the
transverse displacement $\cvs$ to the source terms in \refeq{eomy}.
This is only formally accurate up to fourth order, since $\cvs$ yields
a trace-free contribution to the symmetric part of $\v{H}$
(see e.g. \cite{2012JCAP...06..021R}). 
These contributions are derived in \refapp{curl}.
We have found that their impact on all statistics considered
here is very small ($\lesssim 10^{-4} - 10^{-3}$ depending on the value of the cutoff), significantly smaller at low $z$ than the effect of
incorporating the exact $\Lambda$CDM expansion history. 
This is because the transverse part of the
displacement is always much smaller than the longitudinal part (note that this
is also a consequence of assuming only scalar initial perturbations, i.e. vanishing initial $\cvs$). 
On the other hand,
the additional shapes generated by source terms involving $\cvs$ substantially
increase the memory requirement.

\begin{table}[b]
\centering
\begin{tabular}{c l}
\hline
\hline \\[-2.6ex]
$n$ & Shapes contributing to $\v{H}^{(n)}$ (schematic) \\
\hline \\[-2.6ex]
1 & $H^{(1)}$ \\
2 & $H^{(1)} H^{(1)}$ \\
3 & $H^{(1)} H^{(2)}$, $H^{(1)} H^{(1)} H^{(1)}$ \\
4 & $H^{(1)} H^{(3,1)}$, $H^{(1)} H^{(3,2)}$,
$H^{(2)} H^{(2)}$, $H^{(1)} H^{(1)} H^{(2)}$ \\
\hline
\hline
\end{tabular}
\caption{Different schematic shapes contributing to $\v{H}^{(n)}$ at a given order $n$ for general expansion histories.}
\label{tab:shapes}
\end{table}

The construction of the Eulerian density field proceeds as follows:
\begin{enumerate}
\item $\sigma^{(1)} = -\delta^{(1)}$ is obtained from a given linear density field on a uniform cubic grid of size $N_g^3$ (the choice of grid size will be discussed below). In this paper, $\delta^{(1)}$ corresponds to the initial conditions used for the N-body simulations. A sharp-$k$ filter on the scale $\Lin=\L$ is applied.
\item $\v{M}$ and $\cvs$ are constructed on the same grid as described above. 
\item $\v{s}$ is constructed by evaluating \refeq{sdecomp} on the grid. Note that $\v{s}^{(n)}(\vk)$ has Fourier-space support up to $k = n\L$.
\item $\v{s}(\vk)$ is copied to a larger grid of size $N_{\rm CIC}^3$ in Fourier space (``CIC grid''), and Fourier-transformed to real space.
\item The displacement is then evaluated at each position $\v{q}$ in the CIC grid, and mass $m_{\rm CIC}=1$ assigned to the Eulerian position $\vx = \vq + \v{s}(\vq)$  using a cloud-in-cell (CIC) assignment scheme on the grid scale.
\end{enumerate}

The use of pseudo-particles and density assignment in the last step ensures
the conservation of mass at machine precision. This is important in order to
avoid spurious noise on large scales, and would not be ensured if one were to
expand \refeq{dLPT} in the displacement directly.
The approach described here is very similar to that taken in initial conditions generators
for N-body simulations (e.g., \cite{crocce/etal:2006,MUSIC}). The main differences are that we go to
higher orders in LPT, include beyond-EdS corrections, and apply a sharp-$k$
cut on the linear density field before the LPT construction.
Recently, Ref.~\cite{2020arXiv201012584R} described a very similar scheme.

The final result is the Eulerian density field $\d(\vx,\tau)$ on a grid of size $N_{\rm CIC}^3$ at a given
time $\tau$, to any desired order in LPT, and for any expansion history.
We next turn to the construction of the fields appearing in the bias expansion
of tracers.

\section{Bias expansion at leading order in derivatives}
\label{sec:bias}

The Lagrangian distortion tensor $\v{H}(\vq,\tau)$ captures all leading
gravitational observables for an observer comoving with matter on the
trajectory labeled by the Lagrangian coordinate $\vq$, as shown in \cite{MSZ}, where
``leading'' refers to leading order in spatial derivatives. 
This means that, as a function of Lagrangian coordinate and at leading order
in derivatives, the galaxy density
can be written as a nonlinear functional in time, $F_g$, of $\v{H}$:
\be
\d_g(\vq,\tau) = \int_0^\tau d\tau'\, F_g[\v{H}(\vq,\tau'); \tau';\tau].
\label{eq:dg1}
\ee
In the perturbative regime, we then expand $F_g$ in powers of $\v{H}$,
leading to the appearance of all rotational invariants of $\v{H}(\vq,\tau')$,
such as $\tr[\v{H}]$, in general evaluated at different points in time.
If every contribution to these invariants can be
written in separable form, as in \refeq{sepsigmat}, then the integral
over $\tau'$ in \refeq{dg1} can be formally done, leading to a bias expansion
of the form
\be
\d_g(\vq,\tau) = \sum_O b_O(\tau) O^L(\vq,\tau),
\label{eq:dg2}
\ee
where $b_O(\tau)$ absorbs the unknown small-scale physics encoded in $F_g$,
and the sum runs over a specific set of Lagrangian bias operators $O^L$.
Note that we have to allow for a different bias coefficient for each distinct
time dependence appearing in the expansion of the invariants constructed out
of $\v{H}$ (see \cite{senatore:2015} for the analogous treatment in the
Eulerian case). While this bias construction was previously largely done
under the simplifying EdS assumption, the approach continues to work for general
expansion histories thanks to the separable LPT solution for $\v{H}$
in \refeq{sepsigmat}.

Let us now describe how to construct the set of complete bias operators $O^L$
at leading order in derivatives. We begin by separating the distortion tensor $\v{H}$ into symmetric and antisymmetric parts:
\be
H_{ij} = M_{ij} + C_{ij}, \quad\mbox{where}\quad
M_{ij} = M_{ji};\quad C_{ij} = -C_{ji}.
\ee
This is also convenient for the numerical implementation. The relation
of $\v{M}$ and $\v{C}$ to the longitudinal and transverse parts of the
displacement, $\sigma$ and $\cvs$, is given in \refeq{MC}.

The set of Lagrangian bias operators $O^L$ now comprises all scalar combinations
(rotational invariants) of the contributions to the \emph{symmetric part} of the Lagrangian distortion tensor, $\v{M}^{(n,p)}$, with the
following exception: $\sigma^{(n,p)} \equiv \tr[\v{M}^{(n,p)}]$ with $n > 1$ and any $p$ does not have to be included, since it can be re-expressed in terms of lower-order terms
using the LPT recursion relations. This implies that the bias expansion at order $n>1$ only requires the $\v{M}^{(m,p)}$ up to $m=n-1$.
The transverse displacement $\cvs$ is likewise determined by the set of $\v{M}^{(m,p)}$
at lower orders, and hence does not need to be included
separately. For this reason, it is sufficient to include only the symmetric
part $\v{M}$ in the bias expansion (whether or not the contribution from
$\cvs$ is included in $\v{M}$ is irrelevant for the completeness of the bias
expansion, and only corresponds to a change of basis).

Previous references (e.g., \cite{biasreview}) have listed the complete set of
bias operators (at leading order in derivatives and assuming the EdS approximation) up to fourth order. We now
describe how the general set of operators is constructed up to any order.
The building blocks are the set $\{\v{M}^{(m,p)}\}^p_{m=1,\ldots n-1}$, where $p$
again denotes the different shapes that exist at a given order $m$ for a
general expansion history; in case of the EdS approximation, there is only
a single shape at each order. The enumeration of the $n$-th order bias
expansion then proceeds as follows:
\begin{enumerate}
\item We first construct all scalar invariants up to including $n$-th order out of the $\v{M}^{(m,p)}$. Given the restriction on $\tr[\v{M}^{(m,p)}]$, and since these are symmetric 3-tensors, the invariants at order $m$ consist of the set
  \ba
 \mathcal{I}^{(m)} = \bigg\{&  \tr[ \v{M}^{(1)} ], \quad
 \{ \tr[ \v{M}^{(m_1,p_1)}\v{M}^{(m_2,p_2)} ] \}^{p_1,p_2}_{m_1+m_2\leq m}, \vs
 & \{ \tr[ \v{M}^{(m_1,p_1)}\v{M}^{(m_2,p_2)}\v{M}^{(m_3,p_3)} ] \}^{p_1,p_2,p_3}_{m_1+m_2+m_3\leq m} \bigg\} \vs
\equiv\;& 
\left\{ I_s^{(m)} \right\}_{s=1}^{N_{\mathcal{I}}(m)} \,.
  \ea
\item We then construct all independent products
  \ba
 & I_{s_1}^{(m_1)} \cdots  I_{s_k}^{(m_k)}, \quad 1 \leq k \leq n ,\vs
 &\mbox{with}\quad m_1 + \cdots + m_k = n; \quad
  s_i \in \{1, \ldots, N_{\mathcal{I}}(m_i)\}.
 \ea
 Technically, this is done iteratively by running over the set of partitions of $n$, and then, for each partition $\{m_i\}_{i=1}^k$, constructing products of all combinations of the $\{ s_1, \ldots s_k\}$.
\end{enumerate}

This construction works up to any order $n$, although the number of independent
operators increases rapidly toward larger $n$. 
As already shown in App. C of \cite{biasreview}, the first bias term that
is \emph{not} present in the EdS approximation appears at fourth
order, where $\tr [\v{M}^{(3)} \v{M}^{(1)}]$ generalizes to
\be
\tr [\v{M}^{(3,1)} \v{M}^{(1)}], \quad \tr [\v{M}^{(3,2)} \v{M}^{(1)}].
\ee
The number of additional bias operators that appear when going beyond the EdS approximation
likewise increases rapidly at orders $n> 4$, as can be inferred from \reftab{shapes}.

After the construction of the set of Lagrangian bias operators, each operator is displaced to Eulerian space using the displacement field $\v{s}(\vq)$
following the description at the end of \refsec{LPT}, where the mass $m_{\rm CIC}$ of each pseudo-particle is now given by the respective operator:
\be
m_{\rm CIC}^{O}(\vq) = O^L(\vq).
\ee
This follows the same treatment as in \cite{schmittfull/etal:2018}, where the
Zel'dovich displacement was used for $\v{s}$. 
Note that, due to the Jacobian of the displacement to Eulerian space, the
displacement transforms each operator according to
\be
O^L(\vq) \to O(\vx) = [1 + \d(\vx)] O^L(\vq[\vx]).
\label{eq:Odisp}
\ee
At leading order in perturbations, this corresponds to the desired result, $O^L(\vq[\vx])$. At higher orders, since the bias expansion contains all combinations of operators, the prefactor $1+\d$ can be absorbed by a redefinition of bias parameters.

\section{Higher-derivative bias}
\label{sec:hderiv}

Having described the construction of the complete bias expansion at any
order in \emph{perturbations}, we next turn to the expansion in \emph{derivatives}. In general, this expansion is controlled by the parameter
$(k R_*)^2$, where $k$ is the wavenumber and $R_*$ is a spatial length scale that is specific to
a given tracer. For halos, this scale is expected and found to be of order
the Lagrangian radius \cite{fujita/etal,Abidi:2018eyd,lazeyras/schmidt},
but for real galaxies it could be substantially larger (e.g., \cite{2007JCAP...10..007C,2014PhRvD..89h3010P,2019JCAP...05..031C}), so that
the ability to include higher-derivative bias operators is potentially important. Even for halos, Ref.~\cite{paper_realspace} found them to be numerically
relevant at sufficiently high order in perturbations.

At $k$-th order in derivatives,
the fundamental ingredients of the bias expansion are all invariants constructed
from $\vn\cdots\vn \v{M}^{(m,p)}$, with up to $k$ derivatives acting on any
$\v{M}^{(m,p)}$ (see Sec. 2.6 of \cite{biasreview}). Notice that $k$ has to be even, since all indices need to be contracted
(there are no preferred directions); the number of derivatives acting on a single instance of $\v{M}^{(m,p)}$ can be odd of course.

The full set of higher-derivative terms is unfortunately cumbersome for
several reasons. First, the memory requirements for constructing the tensor $\vn\cdots\vn \v{M}^{(m,p)}$ increase exponentially with the number of derivatives.
Second, there are degeneracies between higher-derivative terms which are not trivial
to remove. As a simple example, consider only terms involving $\sigma^{(1)} = \tr[\v{M}^{(1)}]$. We have, for $k=2$ and $n=2$,
  \be
\lapl( \sigma^{(1)} )^2,\quad (\vn \sigma^{(1)})^2,\quad \sigma^{(1)} \lapl \sigma^{(1)},
\ee
only two of which are independent. In this case, the degeneracy is obvious, but it is
much more difficult to identify and remove at higher orders.

Since an exact degeneracy between different bias terms can lead to severe numerical
issues when sampling parameters or attempting to find the maximum of the likelihood,
we instead opt for the following simplification.
Let $\O$ be the set of Eulerian bias operators constructed up to some
order $n$ in perturbations and $k$ in derivatives, and $n(O)$ be the order
in perturbations of a given operator $O$. The set at order $n$ in perturbations and $k+2$ in derivatives is then constructed by adding the set
\be
\{ \lapl O \}_{O\in\O} \cup \{ \vn O' \cdot\vn O \}_{O,O'\in\O:\, n(O)+n(O') \leq n}
\cup \{ O' \lapl O \}_{O,O' \in \O:\, O\neq O' \wedge  n(O)+n(O') \leq n} 
\ee
to the set of operators, where for the terms involving two operators $O,O'$ only one permutation of each pair is included. This construction is used to iteratively add derivatives
starting from $k=0$ to the desired order.

While this approach only includes a subset of all \emph{nonlinear} higher-derivative bias operators,
these are guaranteed to be linearly independent, and no significant additional
memory is needed (apart from that needed to store the operators themselves). The leading higher-derivative terms that are not included are
\be
\partial_k M^{(1)}_{ij} \partial^k M^{(1)ij}
\ee
and other contractions of the same type, while all higher-derivative terms involving
$\tr[\v{M}^{(1)}]$ are included.

In practice, since the displacement from Lagrangian to Eulerian space
is the costliest step in the forward model, we generate the higher-derivative
operators in Eulerian space, i.e. apply $\vn = \vn_x$ after the displacement
operation \refeq{Odisp}. For a complete set of higher-derivative operators,
the choice of Eulerian or Lagrangian derivatives is irrelevant, as the
differences between $\vn_q$ and $\vn_x$ can be absorbed by shifting
bias parameters, via \refeq{Mdef} (see Sec. 2.6 of \cite{biasreview}). 
For the reasons explained above, the set of higher-derivative operators
implemented here is not complete, so the corresponding Lagrangian and
Eulerian subsets are not equivalent. We leave a detailed investigation
of the practical relevance of the missing higher-derivative operators, and the
choice of Eulerian vs Lagrangian derivatives to future work.

\section{Numerical implementation}
\label{sec:code}

The forward model described above is implemented in an OpenMP-parallelized
\CC\ code centered around classes encapsulating scalar, vector, and tensor
grids. Fast Fourier transforms of the grids are done using FFTW3.
Multiplications are performed in real space, while derivatives (including
negative powers) are performed in Fourier space. The grid resolution is always
chosen to capture all relevant modes, so that no kernel deconvolutions are
necessary.
These aspects are very similar to the implementation described in
\cite{2020arXiv201012584R}, which however is distributed-memory capable.

Several density assignment schemes are available, including nearest grid
point and Fourier-Taylor up to next-to-leading order (see \cite{paperIIb}).
However, all results shown here use the standard cloud-in-cell density
assignment on the given grid scale; specifically, we choose $N_{\rm CIC}=512$
for the Eulerian assignment grid in the forward model (see \refsec{LPT}), as well
as the simulation output (dark matter particles and halos). The
Nyquist frequency of the Eulerian grid for the simulation box used here,
$\Lbox = 2000\Mpch$ is
$k_{\rm Ny}^{\rm CIC} \simeq 0.8\iMpch$, far beyond the maximum wavenumbers
explored here. Moreover, by using the same assignment scheme in the forward
model as well as for simulated tracers, the assignment kernel cancels in the comparison.

The code additionally contains functionality to copy grids to higher and
lower resolutions, matching all Fourier modes up to the smallest Nyquist
frequency involved. We refer to \cite{paper_realspace} for the details
of the Nyquist plane treatment.
A public release of the code is planned for the future.

Finally let us turn to the choice of grid resolution for the forward
model. Throughout, we consider an $n$-th order forward model constructed from
the linear density field with a sharp-$k$ filter with cutoff $\L$.
The results of this forward model are then only used in comparison to
``observations'' on scales $k < \L$. Then, 
there are two possible choices:
\begin{enumerate}
\item \emph{No mode left behind:} in order to ensure that \emph{all} couplings of the
  linear modes up to $k=\L$ are captured, one has to choose $k_{\rm Ny} = n \Lambda$, which implies
  \be
N_g(\L) = \left\lceil n \frac{\Lbox \L}{\pi}\right\rceil.
\label{eq:NgNMLB}
\ee
\item \emph{Generalized Orszag rule:} if the only concern is to avoid aliasing of high-$k$ modes to low $k$, it is sufficient to require
  \be
N_g(\L) = \left\lceil \frac{n+1}2 \frac{\Lbox \L}{\pi}\right\rceil .
\label{eq:NgOrszag}
\ee
This is because the lowest aliased modes have wavenumbers $k_{\rm alias} = 2k_{\rm Ny} - n \Lambda$, which has to be greater than $\Lambda$ to avoid aliased
modes on the range of wavenumbers used (see also \cite{baldauf/LPT}; the special case of $n=2$ is also known as Orszag's rule \cite{Orszag}).
\end{enumerate}
The memory requirements of the first option are significantly higher.
If the mode coupling of high-$k$ modes far above the cutoff $\L$ has
a subdominant effect on the low-$k$ modes used in the comparison with data,
then the second option is a good approximation. In fact, in the effective
field theory (EFT) approach the effect of this type of mode coupling is absorbed by the
counterterms; hence, the generalized Orszag rule might be sufficient if
counterterms are allowed for. We will present some results in this scheme
below, since it allows for higher cutoff values to be probed, but leave
a more detailed exploration (including the possibility of further reducing
$N_g$) for future work.

The reasoning above does not apply to the Eulerian grid of size $N_{\rm CIC}^3$
used for the density assignment. This is a fully nonlinear (in terms of the relation between displacement and resulting field), but local operation. No
aliasing is induced for low-$k$ modes, since the procedure preserves mass
to machine precision. Moreover, the same assignment procedure is applied to
the data (N-body particles in case of the results here; halos in case of
\cite{paper_realspace}).

The simulation results used here are the same as those used in \cite{paperII,paperIIb,paper_realspace}, performed using the GADGET-2 code \cite{2005MNRAS.364.1105S}. 
Specifically, we use simulations started with particles displaced using 2LPT at the starting redshift $z_{\rm in}=24$, as these have substantially reduced transients
(see \cite{paper_realspace} for a discussion). We also make use of
simulations initialized in the same way, but using a linear density field with a
sharp-$k$ cutoff at $\Lin = 0.1\iMpch$ and $\Lin=0.2\iMpch$. 

\section{Comparison with N-body simulations}
\label{sec:sim}

We now turn to the comparison of the forward model described here with
full N-body simulations. We focus here on the matter density field itself.
Results for dark matter halos, specifically cosmology inference using
the EFT likelihood, were already shown in \cite{paper_realspace}.
We leave further studies on halos and other tracers, such as comparisons
of $n$-point correlation functions, to future work.

In particular, we will consider three comparisons here: \emph{(1) Matter power
spectrum in simulations with cutoff:} in this case, the N-body code essentially solves LPT to arbitrarily high order, so that this comparison serves to test the LPT implementation as well as the asymptotic series of the LPT expansion
(see also \cite{coles/etal:1993}).
\emph{(2) Matter power spectrum and correlation coefficient in simulations without cutoff:} the comparison with full, non-perturbative simulations allows for the estimation of the size and scale dependence of the EFT counterterms for matter, in particular the effective sound speed.
\emph{(3) $\sigma_8$ profile likelihood:} as a test for the quality of the
LPT prediction for higher-order statistics, we study the $\sigma_8$ inference
from dark matter particles allowing for the linear bias $b_1$ to be free,
which removes the linear-order information on $\sigma_8$ in the matter
density field (corresponding essentially to the power spectrum).

\subsection{Simulations with cutoff}

\begin{figure*}[htbp]%
  \centerline{\resizebox{\hsize}{!}{
      \includegraphics*{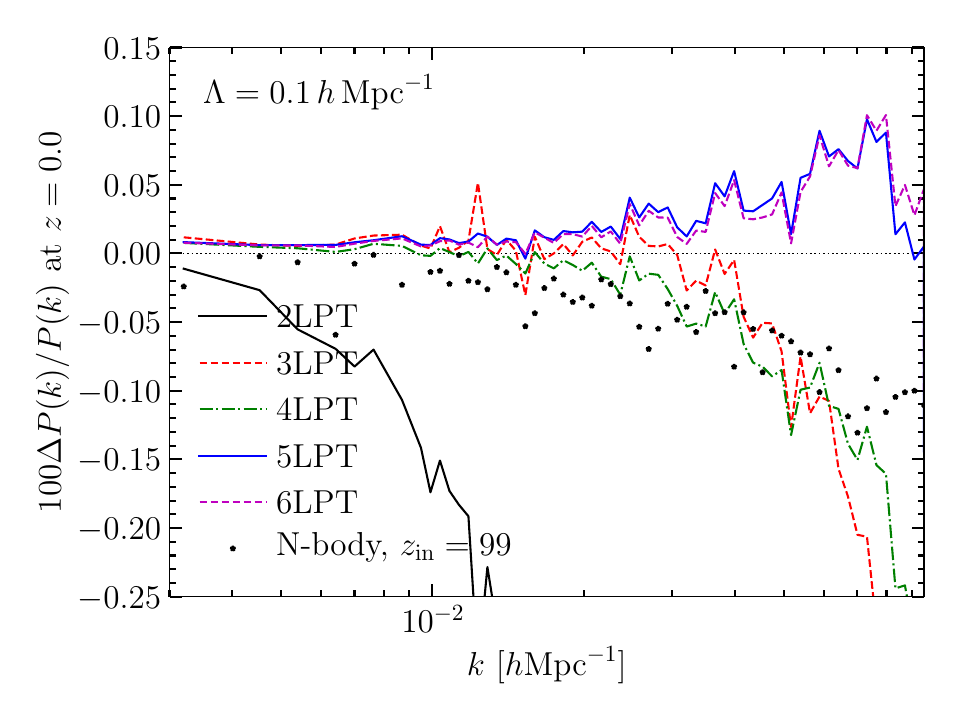}
      \includegraphics*{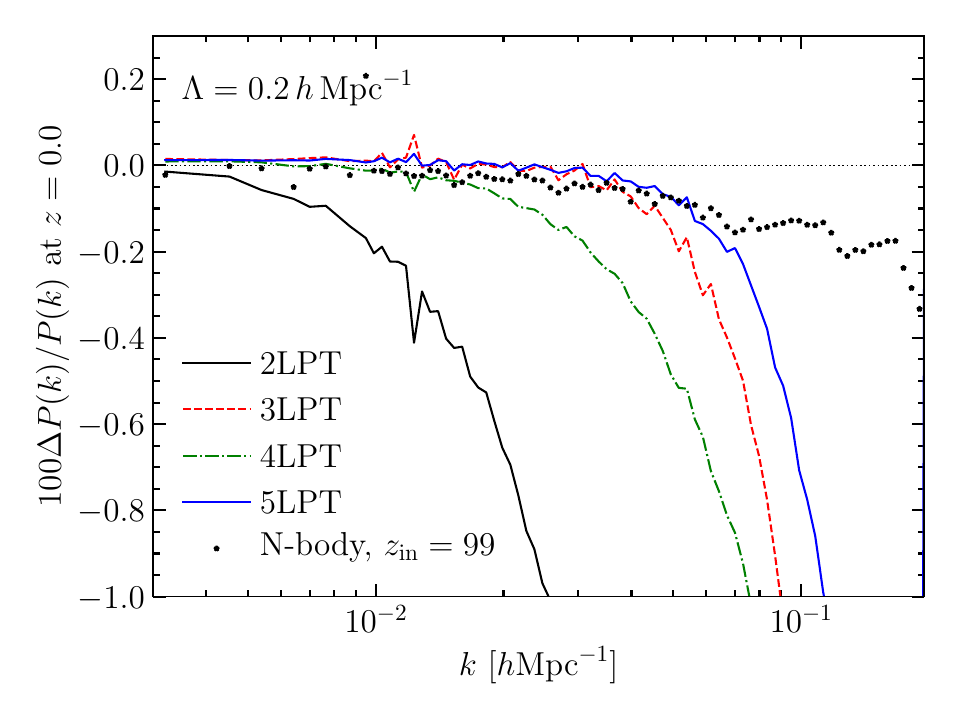}
      }}
  \centerline{\resizebox{\hsize}{!}{
      \includegraphics*{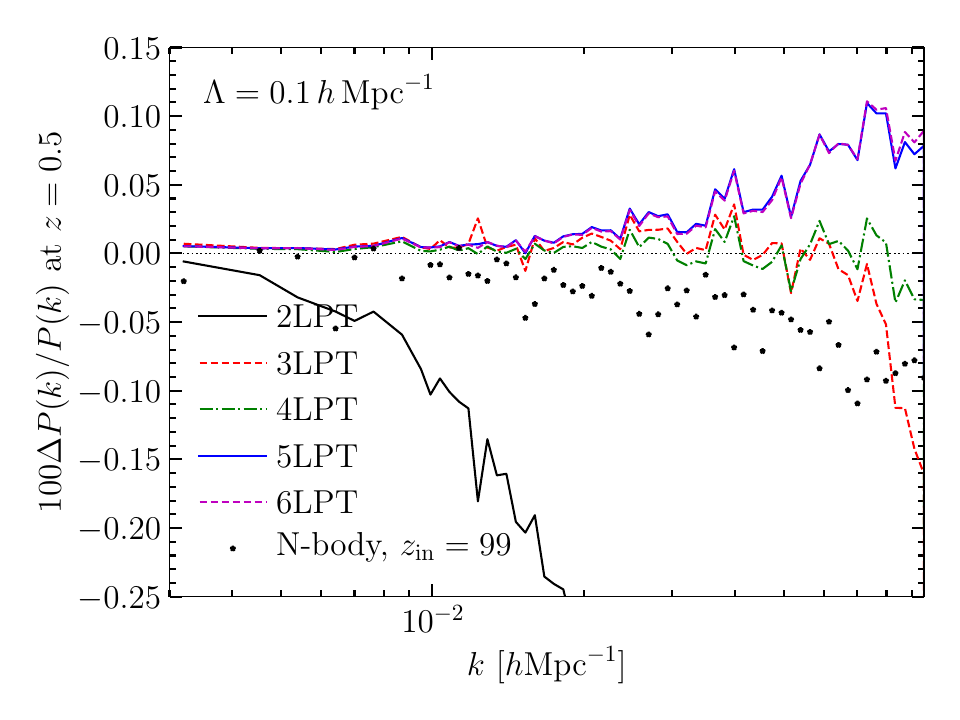}
      \includegraphics*{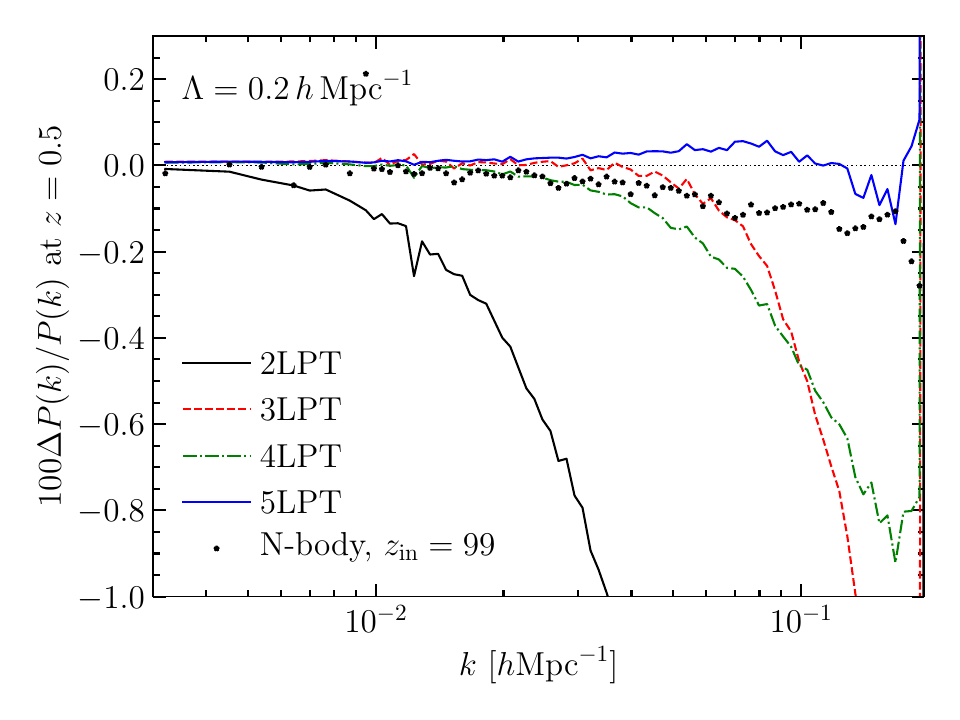}
      }}
  \centerline{\resizebox{\hsize}{!}{
      \includegraphics*{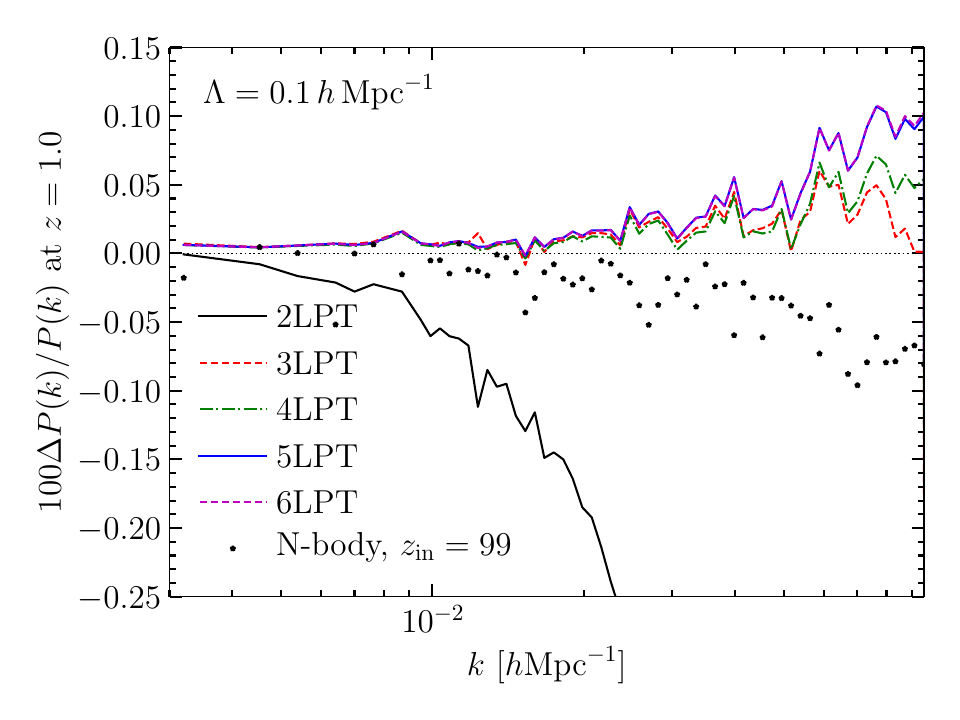}
      \includegraphics*{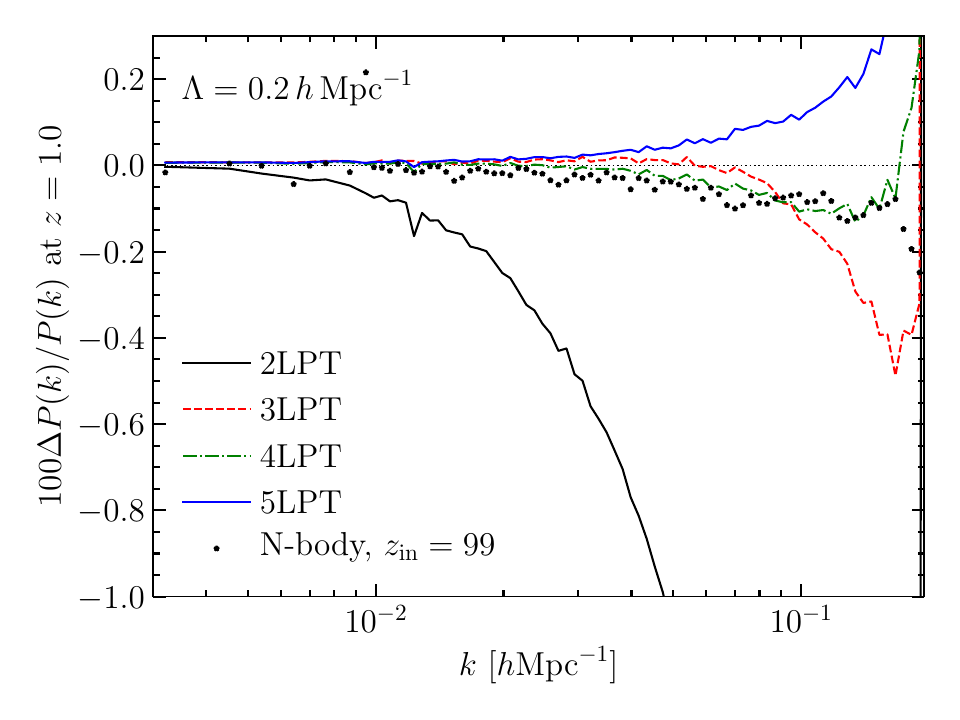}
      }}
  \cprotect\caption{Relative deviation (in percent) of the matter power spectrum measured in the
    LPT forward model output at different orders from that measured in N-body simulations, both using the same initial cutoff $\L$, Eulerian density grid, and assignment kernel. The left panels show results for $\L=0.1\iMpch$, while the right panels show the same for $\L=0.2\iMpch$, each at redshifts $z=0, 0.5, 1$ (top, middle, and bottom panels, respectively). The exact time dependence for each shape is used in the LPT forward model. The dotted points show the corresponding deviation of the matter power spectrum in an N-body simulation started at $z_{\rm in}=99$ instead of $z_{\rm in}=24$, illustrating the order of magnitude of transients from the initial conditions.
    \label{fig:Pk_nLPT_general}}
  \vspace*{0.2cm}
\end{figure*}%

\reffig{Pk_nLPT_general} shows the fractional deviation (in percent) between the power
spectrum in the LPT forward model and that in N-body simulations that use
the same cutoff in the initial conditions. This very large-scale cutoff,
by simulation standards, essentially makes the result of the N-body code
an arbitrary-order LPT solution, so that the asymptotic properties of the
LPT series expansion can be tested. It is worth emphasizing that most of the
cases shown in \reffig{Pk_nLPT_general} are likely beyond the strict
convergence radius of LPT; see \cite{2020arXiv201012584R} for a recent numerical determination
of this scale using a $\Lambda$CDM linear power spectrum, and
\cite{zheligovsky/frisch,rampf/villone/frisch} for more theoretical considerations.
Thus, one expects the deviation of successively higher LPT orders to
shrink initially, and then increase again as higher-order terms are no
longer suppressed.

Even keeping this in mind, one should note that for $n>2$, $n$-th order LPT agrees
with the N-body result to better than $\sim 0.1\%$ for $k \lesssim \L$
in the case of $\L=0.1\iMpch$. This excellent agreement is made possible
by using precisely the same density assignment scheme for both LPT as
well as N-body ``particles,'' and by using the ``no mode left behind''
requirement for the grid size. The agreement with simulations is noticeably
worse if the generalized Orszag rule is used instead, showing that,
when comparing to simulations with a cutoff, the coupling of modes significantly
above the cutoff to lower-$k$ modes is not negligible. 

At $\L=0.1\iMpch$ and $z=0$, one can conclude that the LPT expansion
improves at
least to order $n=6$. Note however that the improvement when going from $n=3,4$
to $n=5,6$ is less clear at higher $z$. 
Turning to the higher cutoff value of $\L=0.2\iMpch$, we see qualitatively
similar results to the lower cutoff case, but with stronger deviations from
the N-body result, as expected. In this case, the large memory requirements
(a factor of 8 times larger than for $\L=0.1\iMpch$ at the same order)
restrict us to fifth order. Even at $z=0$, 5LPT still significantly improves
upon 4LPT.

The main caveat in this comparison is that the N-body simulations necessarily have transients
from the initial conditions, which become more important at higher redshifts.
The points in \reffig{Pk_nLPT_general} show the deviation in the matter
power spectrum in simulations started from 2LPT initial conditions at
$z_{\rm in}=99$ instead of the default $z_{\rm in}=24$, with identical
parameters otherwise. For $\L=0.1\iMpch$, these deviations, which are
thus due to transients from the initial conditions, are of the same
order as the difference between the simulations and high-order LPT.
This also holds for $\L=0.2\iMpch$ at $z=1$. 
Hence, firm conclusions on the asymptotic properties and accuracy
of LPT in these cases can only be reached once transients have been
rigorously quantified. We will defer this study to future work; see e.g. \cite{mccullagh/etal,Nishimichi:2018etk,Michaux:2020yis} for in-depth discussions of transients.

\begin{figure*}[htbp]%
  \centerline{\resizebox{\hsize}{!}{
      \includegraphics*{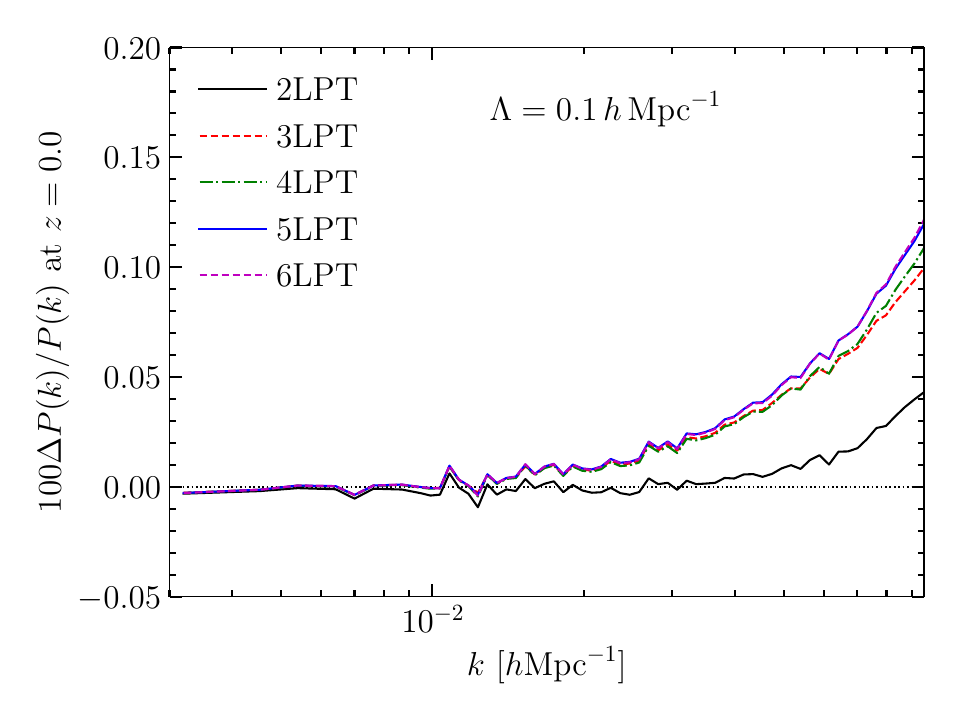}
      \includegraphics*{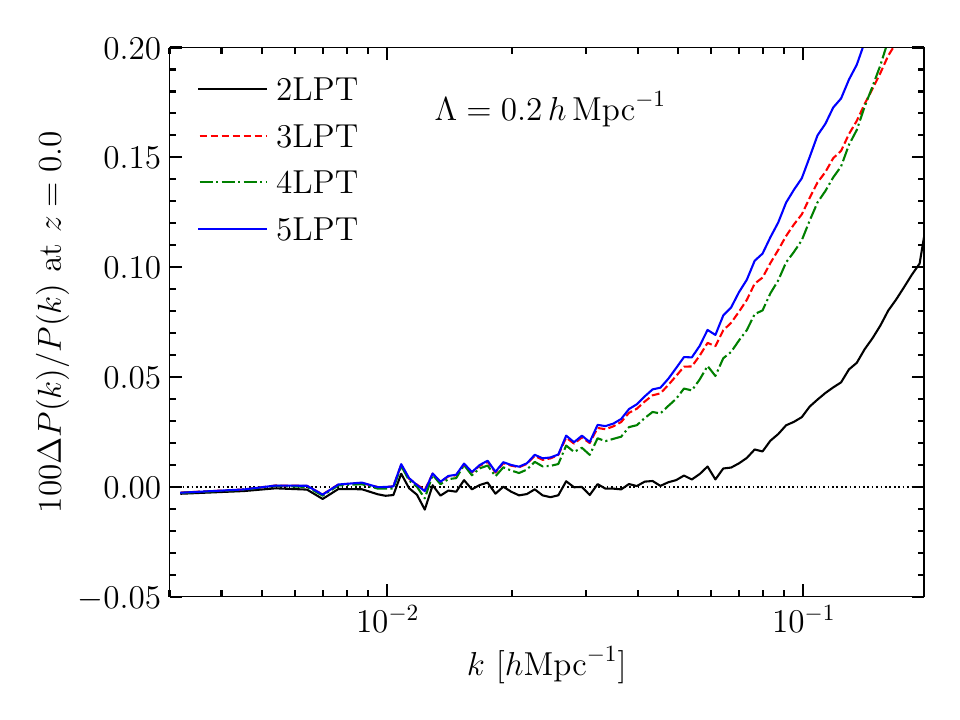}
      }}
  \cprotect\caption{Ratio of LPT power spectra for the general expansion history and using the EdS approximation, both with initial cutoff at $\Lin=0.1 \iMpch$ (left panels) and $\Lin=0.2\iMpch$ (right panels) at $z=0$. In case of $\Lambda$CDM, the prediction using the exact expansion history is always slightly
    larger than the result using the EdS approximation.
    \label{fig:Pk_nLPT_EdS}}
\end{figure*}

The results shown so far use the general time dependence of each shape
by numerically integrating the corresponding equations in $\lambda=\ln D$ for
the $\Lambda$CDM expansion history. However, the change from the EdS approximation is minor, as shown in \reffig{Pk_nLPT_EdS} for $z=0$, where the effects of the
different expansion histories is largest.  The predictions for the exact
expansion history are larger than the EdS approximation by $0.1-0.2\%$,
depending on the cutoff value. This does generally
improve the agreement with the N-body result noticeably. While a correction
of order $0.1-0.2\%$ might be negligible in many applications, it is
worth noting that the magnitude of this correction will depend on cosmology.
In particular, a cosmological model that differs from EdS more substantially
than $\Lambda$CDM, especially at higher redshifts, might lead to bigger
effects. 
Unfortunately, due to the greatly increased number of shapes (grids) that
need to be independently computed, the memory requirements for the general
expansion history case are significantly larger than for EdS.

\subsection{Full simulations}

\begin{figure*}[htbp]%
  \centerline{\resizebox{\hsize}{!}{
      \includegraphics*{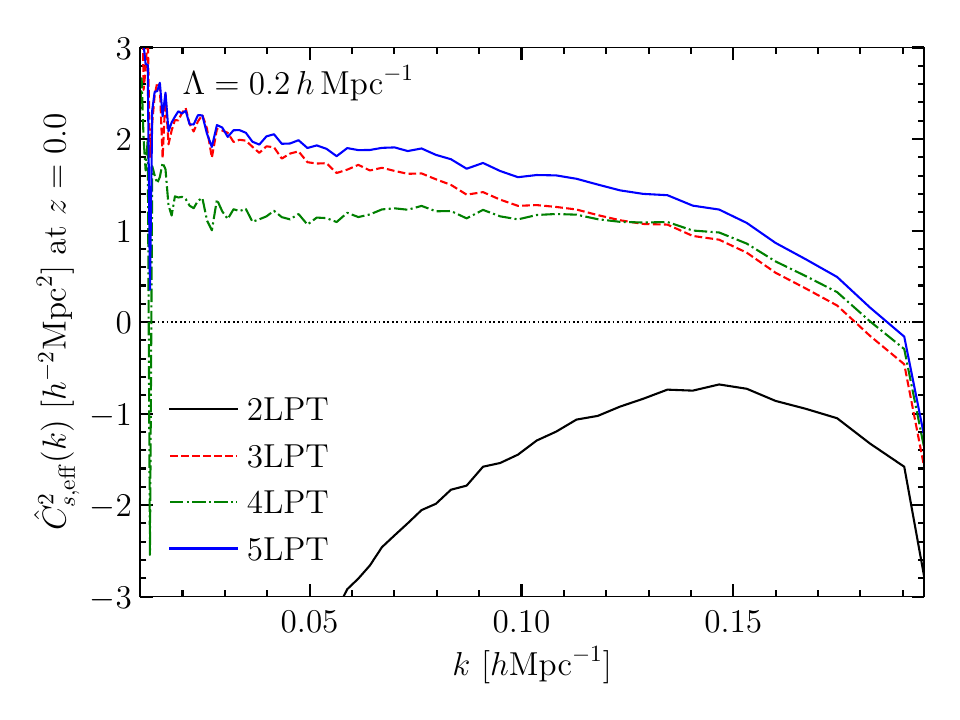}
      \includegraphics*{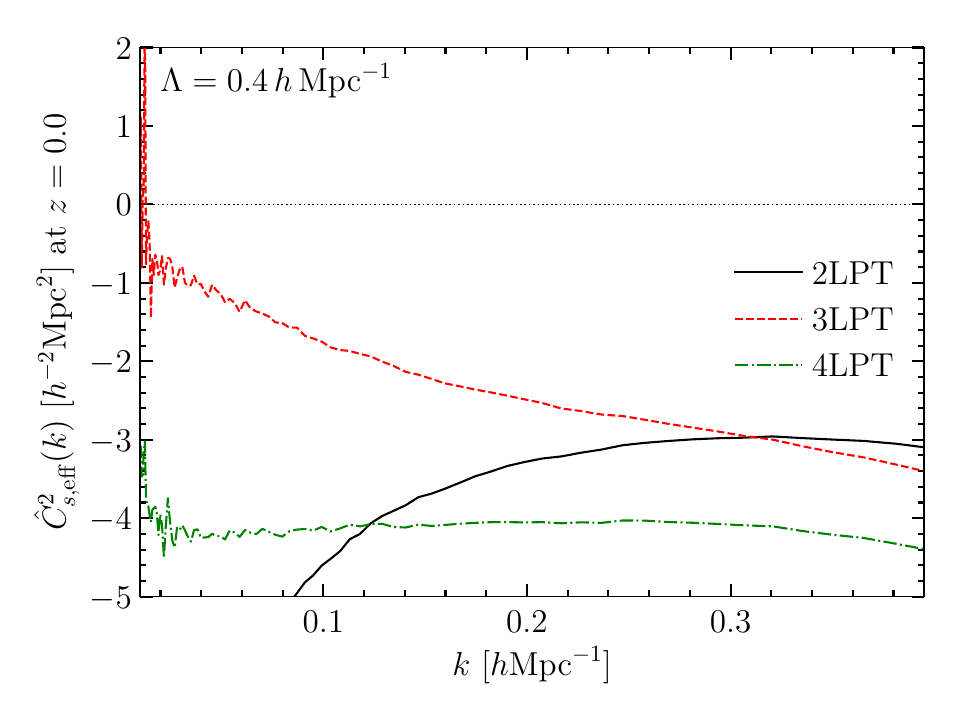}
      }}
  \centerline{\resizebox{\hsize}{!}{
      \includegraphics*{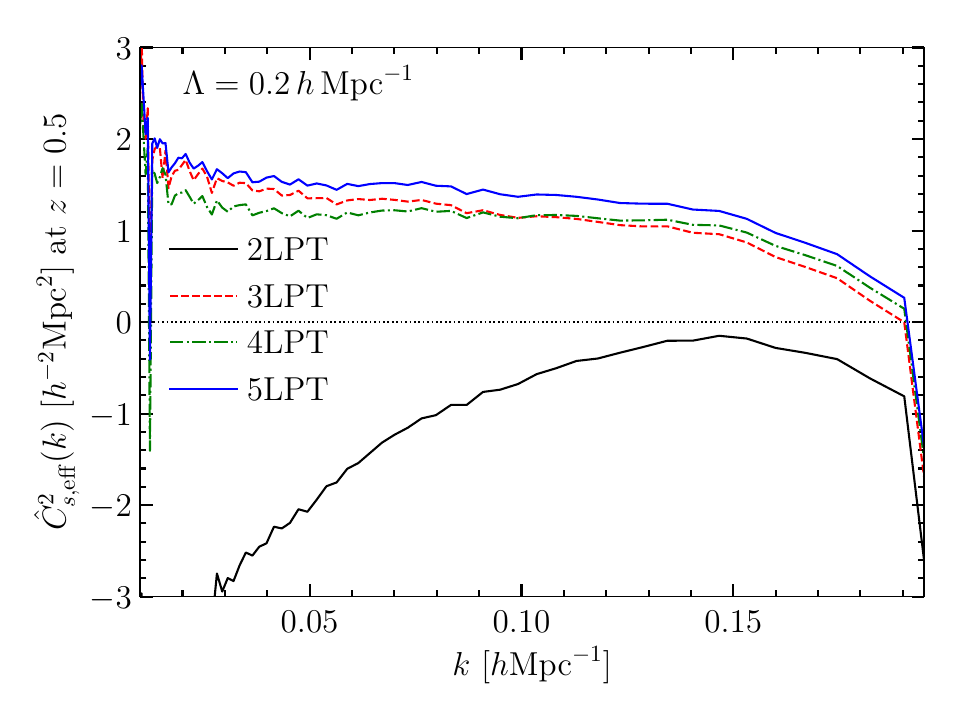}
      \includegraphics*{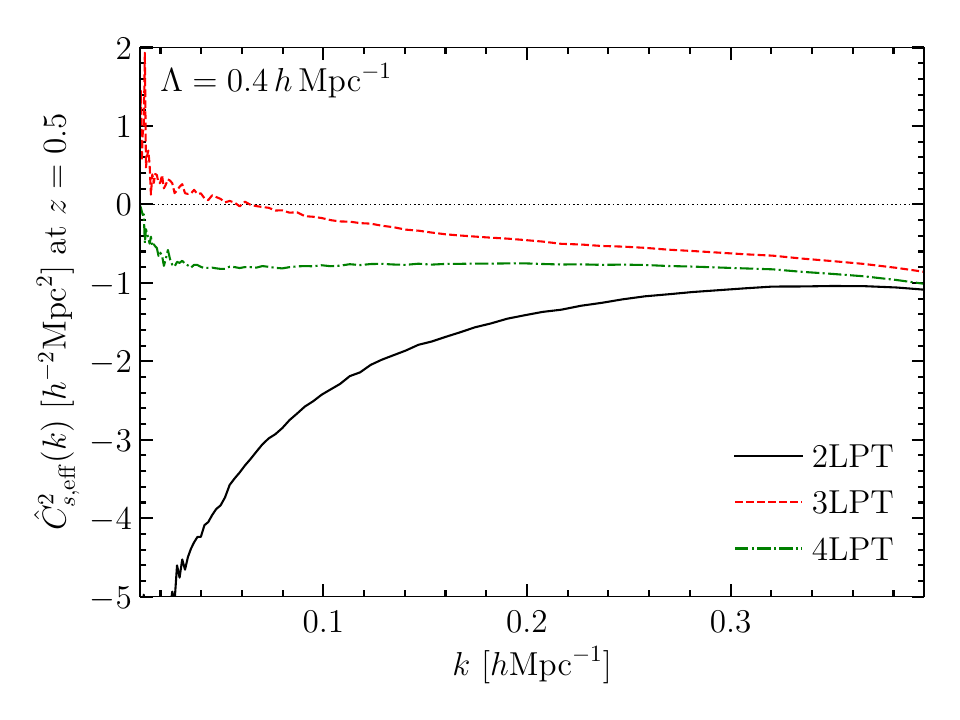}
      }}
  \centerline{\resizebox{\hsize}{!}{
      \includegraphics*{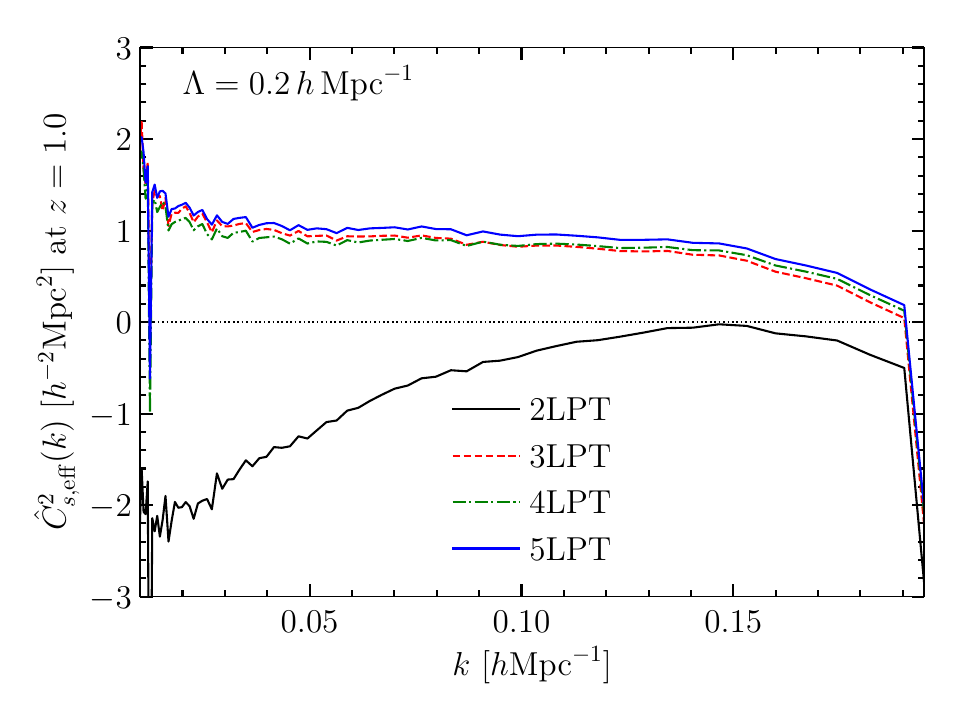}
      \includegraphics*{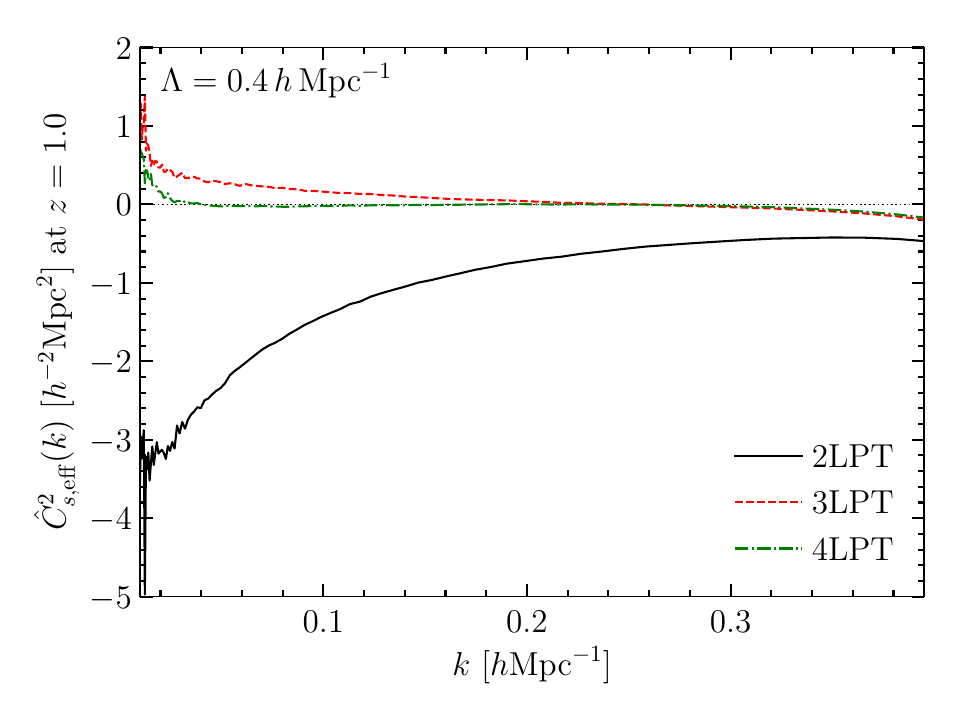}
      }}
  \cprotect\caption{Effective sound speed of matter $\hat C_{s,\rm eff}^2(k)$
    for full N-body simulations,
    estimated through \refeq{Cshat} 
    for different LPT orders and  cutoff values of $\L=0.2\iMpch$ (left panels) and $\L=0.4\iMpch$ (right panels), at different redshifts $z=0,0.5,1$
    (top, middle, and bottom panels, respectively). The exact time dependence is used in the LPT construction. For the cutoff value of $\L=0.4\iMpch$,
    we use the generalized Orszag rule for determining the minimum $k_{\rm Ny}$ at each order.
    \label{fig:Cseff_nLPT_general}}
  \vspace*{1cm}
\end{figure*}%

We now turn to the comparison with full N-body simulations. In this case,
we expect a residual contribution to the density field from the small-scale,
fully nonlinear modes even for $k<\Lambda$, which at leading order
takes the form of an effective sound speed \cite{baumann/etal:2012,carrasco/etal:2012}:
\be
\d_{\rm c.t.}^{(1)}(\vk,\tau) = - C_{s,\rm eff}^2(\Lambda,\tau) k^2 \d(\vk,\tau).
\label{eq:Cseff}
\ee
Note that for the present application, $C_{s,\rm eff}^2$ also depends on
the cutoff $\L$, although we will not write this dependence explicitly in
what follows for the sake of clarity.
The effective sound speed is only the leading EFT counterterm to the matter density field;
one also expects quadratic contributions as well as a stochastic
contribution to the power spectrum that scales as $k^4$. Here, however,
we will restrict to the investigation of the leading counterterm;
see \cite{baldauf/LPT} for a detailed study of counterterms in the LPT
context.

The counterterm in \refeq{Cseff} contributes to any perturbative
prediction for the matter power spectrum through
\be
P_{m,\rm EFT}(k,\tau) = \left[1 - 2 C_{s,\rm eff}^2(\tau) k^2 \right] P_{m,\rm LPT}(k,\tau) + \mbox{higher order},
\label{eq:Pmeff}
\ee
where we have defined $C_{s,\rm eff}^2$ to multiply the full PT-predicted
density field. Setting the left-hand side equal to the N-body power
spectrum, we can solve for $C_{s,\rm eff}^2$ as a function of scale and time,
\be
\hat C_{s,\rm eff}^2(k,\tau) = - \frac{P_{m,\rm N-body}(k,\tau) - P_{m,\rm LPT}(k,\tau)}{2 k^2 P_{m,\rm LPT}(k,\tau)},
\label{eq:Cshat}
\ee
in our case for different
LPT orders, $P_{m,\rm LPT}$, and for different cutoff values.
At leading order, we then expect a scale-independent result for $\hat C_{s,\rm eff}^2$, while any nontrivial $k$-dependence indicates the presence of higher-order perturbative or counterterm corrections.

The result is shown in \reffig{Cseff_nLPT_general}. The left panels
show a cutoff value which we have seen is still under perturbative control,
i.e. all modes entering the forward model are still in the quasilinear
regime. The trends seen qualitatively reflect those of \reffig{Pk_nLPT_general}: for 2LPT, $\hat C_{s,\rm eff}^2$ shows a strong scale dependence, while the
scale dependence
becomes much weaker at higher orders. 4LPT shows the weakest scale dependence in this case. Note that as $k$ approaches $\L$, the scale dependence of $\hat C_{s,\rm eff}$ becomes stronger. This is expected, since $\hat C_{s,\rm eff}^2$
has to capture the effect of all modes above $\L$. For $k$ approaching $\L$,
the coupling with modes above the cutoff can no longer be captured by the
simple counterterm in \refeq{Cseff}; equivalently, one can think of
\refeq{Cseff} as being an expansion in $k^2/\L^2$, with only the leading
term being included.

Perturbation-theory predictions of the matter power spectrum and other
statistics typically do not employ an explicit cutoff as done in this
forward model (but see \cite{carroll/etal,cabass/schmidt:2019,paperIIb}).
Instead, the cutoff is sent to infinity, or at least large values, so that loop integrals run over
arbitrarily large momenta. In order to get closer to this case, we show
results for $\L=0.4\iMpch$ in the right panels of \reffig{Cseff_nLPT_general}.
In this case, we use the generalized Orszag rule to determine the grid size,
substantially reducing the memory requirements. This can be justified by the
fact that the coupling of several high-$k$ modes to a low $k$-mode that
is being ``left behind,'' i.e. not captured on the grid, should be absorbed
by the counterterm. Interestingly, we indeed find that the scale dependence
of $\hat C_{s,\rm eff}^2$ is moderate for 3LPT, and in fact very small for
4LPT. This indicates that 4LPT with a high cutoff, combined with a single
scale-independent $C_{s,\rm eff}^2(\tau)$, can describe the matter power spectrum
to the few-percent level at least up to $k\lesssim 0.4\iMpch$. Care
should be taken however, as this amounts to extrapolating the perturbative
prediction past its regime of validity.

Several measurements of the effective sound speed have been presented in the
literature \cite{Carrasco:2013sva,Angulo:2015,baldauf/LPT,baldauf/etal:2015,lazeyras/schmidt}, most of which have been based on Eulerian perturbation theory.
The closest comparison point is offered by \cite{baldauf/LPT}. Fig.~9
there shows results for 3LPT and $\L=0.61\Mpch$, measured using different
approaches. The value of $\hat C_{s,\rm eff}^2$ (corresponding to $\alpha$
there) is in broad agreement at $k\simeq 0.1\iMpch$ for both $z=0$
and $z=0.5$. However, the scale dependence shows some differences, which
is likely attributable to the different cutoff $\L$, as well as
different estimators used.

One particular aspect of the results in \reffig{Cseff_nLPT_general} is
noteworthy: while $\hat C_{s,\rm eff}^2$ is in general scale-dependent,
this scale dependence is smooth and does not show any signatures of the BAO.
This means that BAO damping is captured very accurately by the LPT
forward model, which is in contrast to Eulerian PT approaches (see Fig.~6 of
\cite{baldauf/etal:2015} for a clear illustration).

\begin{figure*}[htbp]%
  \centerline{\resizebox{\hsize}{!}{
      \includegraphics*{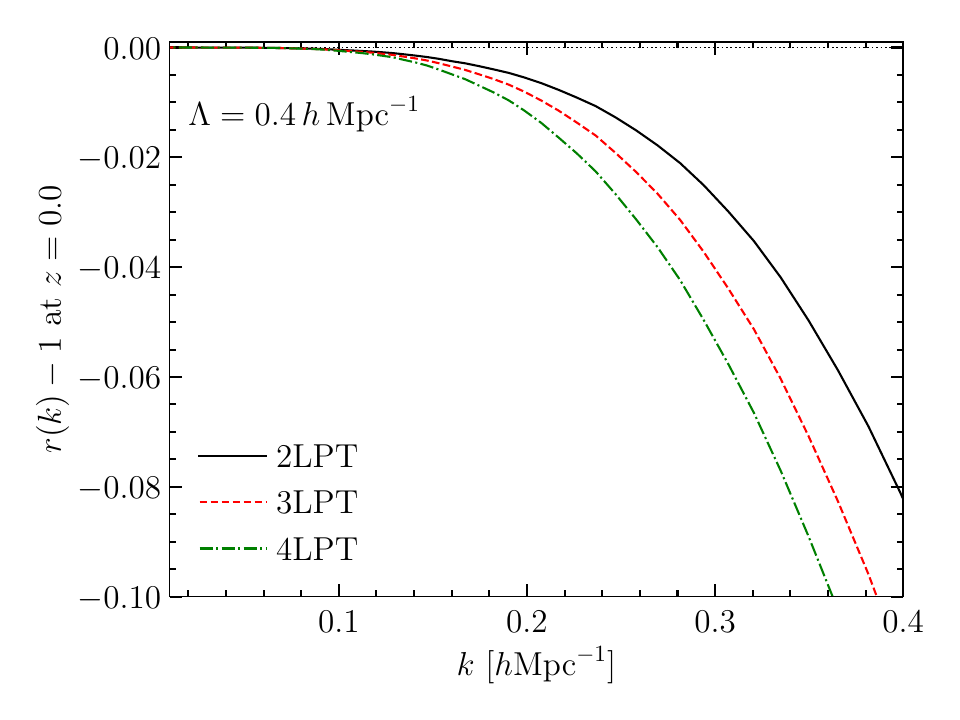}
      \includegraphics*{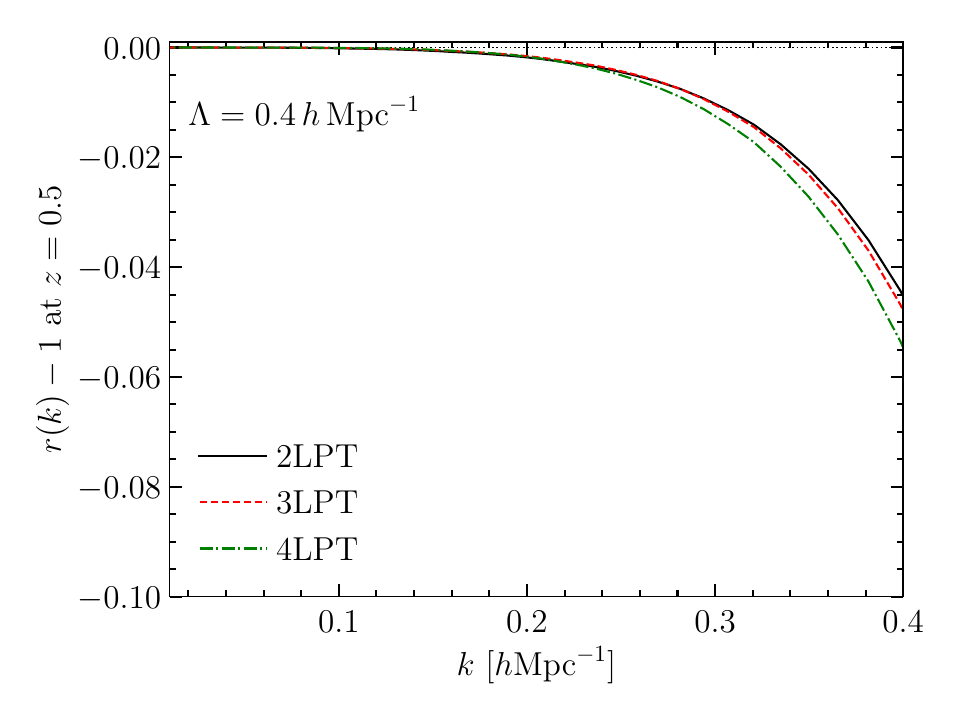}
      }}
  \centerline{\resizebox{0.5\hsize}{!}{
      \includegraphics*{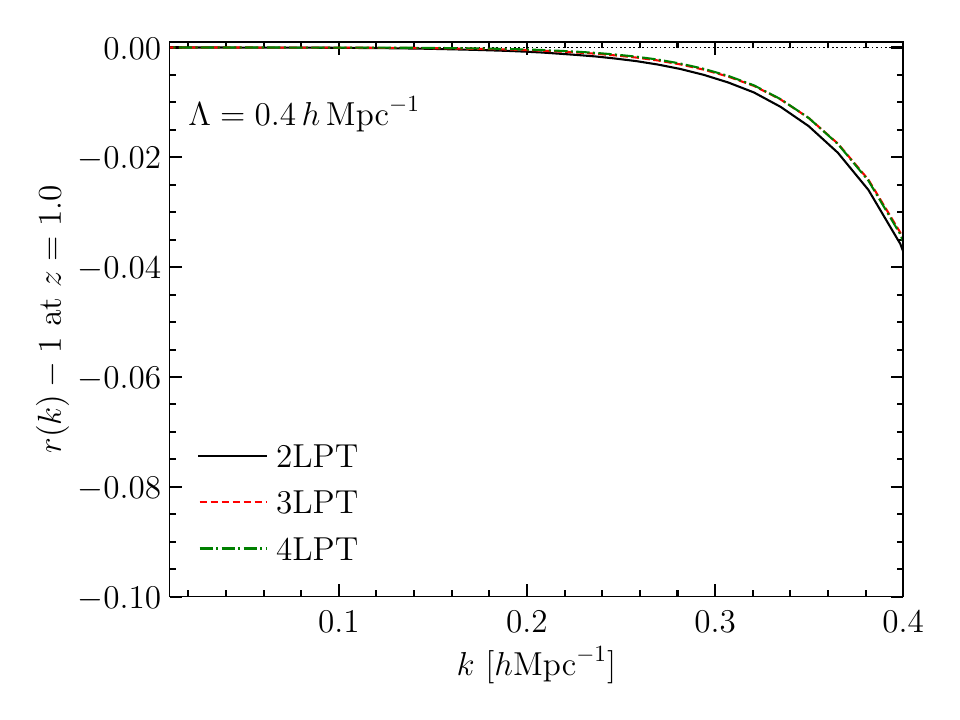}
      }}
  \cprotect\caption{Correlation coefficient between the $n$-LPT density field
    with exact time dependence 
    and the full N-body density field, for $\L=0.4\iMpch$ at different redshifts. Here, we have used the generalized Orszag rule for determining the minimum $k_{\rm Ny}$.
    \label{fig:rk_nLPT_general}}
\end{figure*}%

Finally, \reffig{rk_nLPT_general} shows the correlation coefficient
(more precisely, its deviation from unity)
between the $n$-LPT density field and the full N-body density field.
The correlation is clearly very high even for nonlinear scales, confirming
previous results (going back to \cite{coles/etal:1993,buchert/etal:1994,melott/etal:1995}). Interestingly, at $z=0$, the best correlation
coefficient is obtained by 2LPT, although the difference between the
different LPT orders is rather small. It is also worth noting that
the LPT density using the EdS approximation performs slightly worse (not shown here). 
At $z >0.5$, all LPT orders essentially
yield the same correlation coefficient. The fact that higher LPT orders
lead to a worsening correlation coefficient for $\L=0.4\iMpch$ at $z=0$
is likely attributable to the fact that one is going significantly beyond
the convergence radius of LPT for
such a high cutoff value. The comparison with $\L=0.2\iMpch$ at $z=0$, which
does not exhibit this behavior, confirms this (not shown here).

\subsection{Profile likelihood for \texorpdfstring{$\sigma_8$}{sigma8}}

The profile likelihood as a tool to study cosmology inference from
nonlinear structure in the simulation setting, i.e. when the initial
conditions are known, has been studied
extensively recently \cite{paperII,paperIIb,paper_realspace}.
Those studies focused on the inference of $\sigma_8$ from halo catalogs
in real space. Due to the perfect degeneracy between the linear bias
$b_1$ and $\sigma_8$ at linear order, this inference is based entirely
on nonlinear information.

Here, we perform an analogous study using the dark matter density field
from full N-body simulations as tracer field. In the inference, we leave
$b_1$ and the higher-derivative bias $b_{\lapl\d}$ free, while fixing all
other bias parameters to zero. $b_1$ is left free to ensure
that we test the nonlinear part of the density field, since the
agreement of LPT and N-body density fields on large, linear scales
was already established using the power spectrum above. The bias
term $b_{\lapl\d}$ essentially corresponds to the effective sound speed
\refeq{Cseff}, and so should be allowed to vary in any case.

\begin{figure*}[htbp]%
  \centerline{\resizebox{\hsize}{!}{
      \includegraphics*{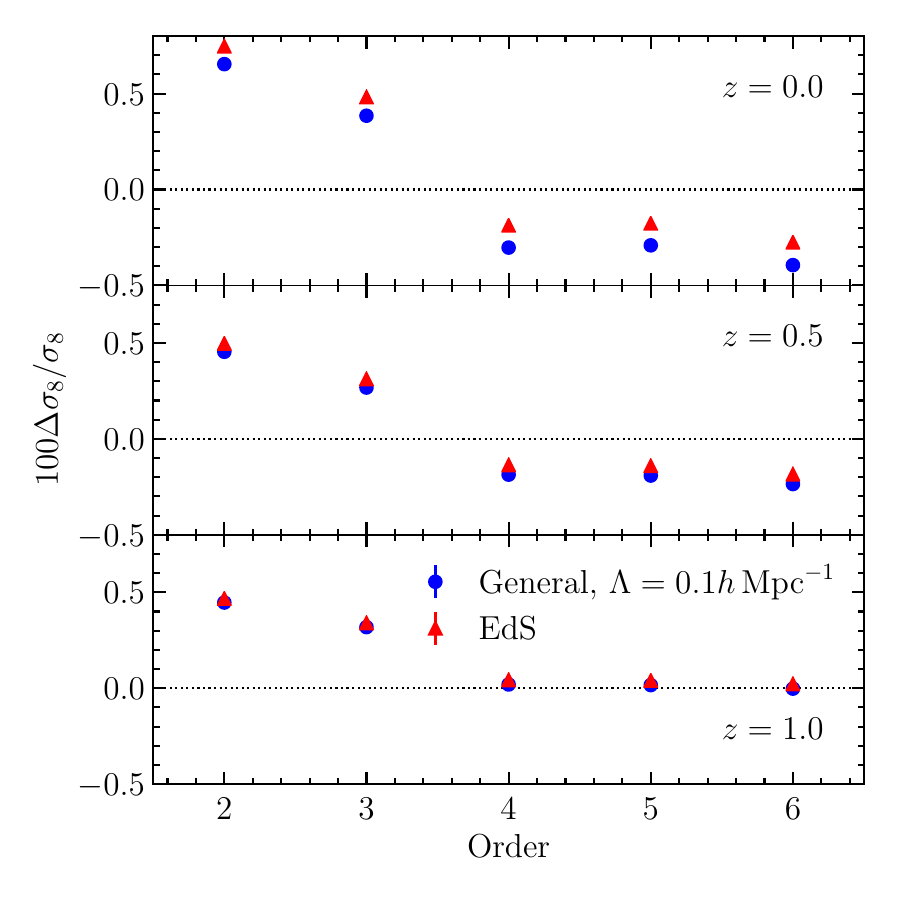}
      \includegraphics*{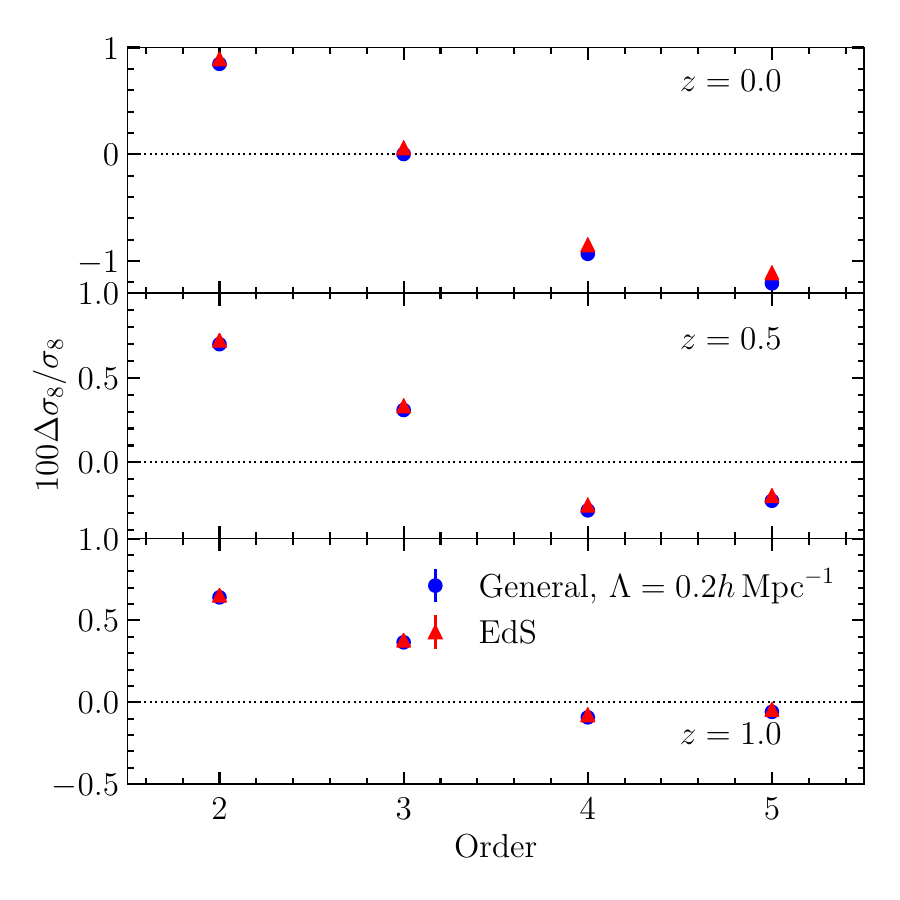}
      }}
  \cprotect\caption{%
    Fractional deviation from the ground truth (in percent) of the maximum-a-posteriori value for $\sigma_8$ inferred from the profile likelihood applied to the N-body matter density field. Results are shown for fixed cutoffs $\L=0.1\iMpch$ (left panels) and $\L=0.2\iMpch$ (right panels) as function of LPT order, at different redshifts $z=0,0.5,1$
    (top, middle, and bottom panels, respectively). In case of $\L=0.2\iMpch$, the statistical error bars are too small to be seen here. 
    Here, the linear bias $b_1$ and higher-derivative bias $b_{\lapl\d}$ were left free, so that $\sigma_8$ inference is based on nonlinear information, while all other bias parameters are fixed to zero. We show results using the exact time dependence as well as the EdS approximation.
    \label{fig:sigma8_vs_order}}
\end{figure*}%

We then proceed exactly as described in \cite{paperIIb} to infer
$\sigma_8$. The result, precisely the fractional deviation in percent of the
inferred value of $\sigma_8$ from the ground truth, is shown in \reffig{sigma8_vs_order}
as function of LPT order. We generally see improved agreement with
increasing order in LPT, as expected in the converging regime of the
asymptotic LPT expansion, with the exception of $\L=0.2\iMpch$ at $z=0$. For $z\geq 0.5$,
LPT at fifth and higher order recovers the correct $\sigma_8$ to within
0.2\% for both cutoff values, where one should recall that this inference is based on
information \emph{beyond the power spectrum} (and for fixed phases). 
This accuracy is very likely to be sufficient
for any application to current and next-generation surveys.
We again find that using the exact time
dependence for $\Lambda$CDM only shifts the inferred $\sigma_8$ by at most $\sim 0.1\%$
relative to the EdS approximation.

\section{Conclusions}
\label{sec:conc}

We have presented an arbitrary-order Lagrangian forward model for the matter
density field as well as that of biased tracers of large-scale structure.
It includes the complete bias expansion at any order in perturbations,
and captures general expansion histories without relying on the EdS approximation (although the latter is also implemented and results in substantially smaller computational demands). 
A subset of the nonlinear higher-derivative terms in the bias expansion of
general tracers is included as well.

We have studied the accuracy of this forward model, and the convergence properties
of the asymptotic LPT expansion, focusing on the matter density field here.
Results for biased tracers (halos) have already been presented in \cite{paper_realspace}.
The findings can be summarized as follows:
\begin{itemize}
\item Comparing the matter power spectrum with N-body simulations using a cutoff $\L=0.1\iMpch$ in the initial conditions, we find agreement to better than 0.1\%.
\item When comparing to full N-body simulations without cutoff, we were able to measure the effective sound speed. For relatively high cutoff values, corresponding to the case commonly employed in perturbative calculations, the effective sound speed is very close to scale-independent all the way up to $k\sim 0.4\iMpch$, with no signs of oscillations, showing that BAO damping is captured very accurately.
\item As a measure of the accuracy of the predicted density field beyond the power spectrum, we studied the inference of the primordial power spectrum normalization $\sigma_8$, allowing for a free linear bias coefficient in order to remove the information from linear perturbations. In the regime where results converge as function of LPT order, we find agreement to 0.2\% or better.
\end{itemize}

It would be interesting to test the convergence of LPT more systematically
in the future, along the lines of  \cite{2020arXiv201012584R}. Simulations
with cutoff can serve as a new tool in this approach. Further, it is
important to test the relevance of the neglected higher-derivative terms
for biased tracers such as halos. Also, the forward model presented here
is uniquely suited to exploring expansion histories beyond $\Lambda$CDM
(e.g. with oscillating equation of state \cite{2017arXiv170901544S}).

Finally, in order to ready the forward model for the application to real data,
redshift-space distortions and light-cone effects need to be included.
Both of these can be straightforwardly incorporated in this approach;
in case of the former, this was explicitly shown by \cite{Cabass:2020jqo},
the latter are described in \cite{2018arXiv180807496K,2019arXiv190906396L}. 
We leave the implementation of these effects for future work.

\acknowledgments

I am indebted to Martin Reinecke for help in optimizing the code,
and to Cornelius Rampf for pointing out the transverse contribution to $\v{M}$.
I further thank Tobias Baldauf, Giovanni Cabass, Oliver Hahn, Donghui Jeong,
Guilhem Lavaux, Cornelius Rampf, Marcel Schmittfull, and Volker Springel for helpful discussions.
Finally, I am grateful to Adrian Hamers for access to his
large shared-memory computer cluster at MPA, which enabled results at higher orders
in LPT. 
I acknowledge support from the Starting Grant (ERC-2015-STG 678652) ``GrInflaGal'' of the European Research Council.

\appendix

\section{Einstein-de Sitter solution of LPT}
\label{app:EdS}

In an Einstein-de Sitter universe, where $\Om(a)=1$ and the linear growth factor is
$D(a) = a$ so that $\gamma\equiv 0$, all shapes of $\v{M}$ at a given order in perturbation theory have the same time dependence $D^n(a)$. Then, the equation of motion \refeq{eomy} can be solved analytically to yield \cite{zheligovsky/frisch,MSZ}
\ba
\sigma^{(n)}(\vq,y) = & \sum_{m_1+m_2=n} c^{(\sigma)}_{n,m_1,m_2}
\bigg\{ \tr\left[ \v{M}^{(m_1)}(\vq,y) \v{M}^{(m_2)}(\vq,y) \right]
- \tr\left[ \v{M}^{(m_1)}(\vq,y) \right]
 \tr\left[\v{M}^{(m_2)}(\vq,y) \right]\bigg\} \vs
& -\frac12 \sum_{m_1+m_2+m_3=n} c^{(\sigma)}_{n,m_1,m_2,m_3} \eps_{ijk}\eps_{lmn} M^{(m_1)}_{il}(\vq,y) M^{(m_2)}_{jm}(\vq,y) M^{(m_3)}_{kn}(\vq,y) \\
 (\cvs^{(n)})^i =& \sum_{m_1+m_2=n} c^{(t)}_{n,m_1,m_2} \eps^{ijk} \left( \v{M}^{(m_1)} \v{M}^{(m_2)\,\top}\right)_{jk}\,,
\ea
where the weights are given by
\ba
c^{(\sigma)}_{n,m_1,m_2} &= \frac12 \frac{m_1^2 + m_2^2 + (n-3)/2}{n^2 + (n-3)/2} ;
\quad
c^{(\sigma)}_{n,m_1,m_2,m_3} = \frac13 \frac{m_1^2 +m_2^2 +m_3^2 + (n-3)/2}{n^2 + (n-3)/2} ; \vs
c^{(t)}_{n,m_1,m_2} &= \frac12 \frac{m_2-m_1}{n}.
\ea
This solution is used as initial condition at $\lambda = \ln(0.1)$ for the integration
of \refeq{eomy} for a general expansion history. Results when
using this solution at all times are also shown in the main text.

\section{Transverse contributions to LPT source terms}
\label{app:curl}

Let us write the Lagrangian distortion tensor at any given order as
\be
H_{ij} = M_{ij} + C_{ij},
\ee
where
\ba
M_{ij} &= M_{ji} = \frac{\partial_i \partial_j}{\lapl} \sigma
- \eps^{(ikl} \frac{\partial_{j)} \partial_k}{\lapl} t_l \vs
C_{ij} &= -C_{ji} = \frac12 \eps_{ijk} t^k
\label{eq:MC}
\ea
are the symmetric and antisymmetric parts of the distortion tensor $\v{H}$,
and we have used the fact that $\cvs$ is divergence-less. 
Note that the transverse or curl part of the displacement also contributes
to the symmetric distortion tensor $\v{M}$. 
In this appendix, all spatial derivatives
are with respect to Lagrangian coordinates $\vq$.

In the code implementation, we add the symmetric transverse contribution
to $\v{M}$ in the first line of \refeq{MC} at each order.
It then remains to derive the contributions of the antisymmetric
part $\v{C}$ to the distortion tensor.

For the longitudinal source, we have, first,
\ba
\tr( \v{H}^{(a)} \v{H}^{(b)} ) &= \tr( \v{M}^{(a)} \v{M}^{(b)} )
+\tr( \v{C}^{(a)} \v{C}^{(b)} ) \vs
&= \tr( \v{M}^{(a)} \v{M}^{(b)} ) + \frac12 \cvs^{(a)}\cdot\cvs^{(b)},
\ea
while $\tr(\v{C})=0$. 
Note that $\tr(\v{M} \v{C})$ vanishes by symmetry as well. 
Second, the transverse contributions to the cubic source term are
\ba
\eps^{ijk} \eps^{lmn} M^{(a)}_{il} M^{(b)}_{jm} M^{(c)}_{kn} \Big|_{\rm transverse}
&=
\eps^{ijk} \eps^{lmn} \hM^{(a)}_{il} C^{(b)}_{jm} C^{(c)}_{kn} + \perm{2}\vs
&= \frac14 \eps^{ijk} \eps^{lmn} \eps_{jmp} \eps_{kno} \hM^{(a)}_{il} t^{(b) p} t^{(c) o} + \perm{2}\vs
&= \frac12 \cvs^{(b) \top}\cdot\v{M}^{(a)}\cdot \cvs^{(c)} + \perm{2}\,.
\ea
Again, the terms involving a single or three transverse components vanish by
symmetry. Since $\cvs$ starts at third order, the corresponding contributions
to the longitudinal source terms start at sixth order.

Turning to the source terms for the transverse component, we have for the transverse contributions
\ba
\eps^{ijk} \left( \v{H}^{(a)}  \v{H}^{(b)\,\top}\right)_{jk} \Big|_{\rm transverse} &=
\eps^{ijk} \left( C^{(a)}_{jm} \hM^{(b)}_{mk} - M^{(a)}_{jm} C^{(b)}_{mk}
- C^{(a)}_{jm} C^{(b)}_{mk} \right) \vs
&= \frac12 \left( t^{(a) i} \tr(\v{M}^{(b)}) - t^{(b) i} \tr(\v{M}^{(a)}) + M^{(a)}_{im} t^{(b) m} - M^{(b)}_{im} t^{(a) m} \right) \vs
&\quad + \frac14 \eps^{ijk} t^{(a)}_j t^{(b)}_k .
\ea
Here, the leading contribution is at fourth order. However, since the transverse part of the displacement is always much smaller than the longitudinal part,
the overall contribution to the source terms remains highly suppressed at
least up to $n=7$.

\bibliographystyle{JHEP}
\bibliography{bibliography}

\providecommand{\href}[2]{#2}\begingroup\raggedright\begin{thebibliography}{10}

\bibitem{paper_realspace}
F.~{Schmidt}, \emph{{Sigma-eight at the percent level: the EFT likelihood in
  real space}},
  \href{https://doi.org/10.1088/1475-7516/2021/04/032}{\emph{\jcap} {\bfseries
  2021} (Apr., 2021) 032}, [\href{https://arxiv.org/abs/2009.14176}{{\ttfamily
  2009.14176}}].

\bibitem{2019MNRAS.482.1786L}
M.~{Lippich}, A.~G. {S{\'a}nchez}, M.~{Colavincenzo}, E.~{Sefusatti},
  P.~{Monaco}, L.~{Blot} et~al., \emph{{Comparing approximate methods for mock
  catalogues and covariance matrices - I. Correlation function}},
  \href{https://doi.org/10.1093/mnras/sty2757}{\emph{\mnras} {\bfseries 482}
  (Jan., 2019) 1786--1806}, [\href{https://arxiv.org/abs/1806.09477}{{\ttfamily
  1806.09477}}].

\bibitem{2012JCAP...04..013T}
S.~{Tassev} and M.~{Zaldarriaga}, \emph{{The mildly non-linear regime of
  structure formation}},
  \href{https://doi.org/10.1088/1475-7516/2012/04/013}{\emph{\jcap} {\bfseries
  2012} (Apr., 2012) 013}, [\href{https://arxiv.org/abs/1109.4939}{{\ttfamily
  1109.4939}}].

\bibitem{baldauf/LPT}
T.~{Baldauf}, E.~{Schaan} and M.~{Zaldarriaga}, \emph{{On the reach of
  perturbative descriptions for dark matter displacement fields}},
  \href{https://doi.org/10.1088/1475-7516/2016/03/017}{\emph{\jcap} {\bfseries
  2016} (Mar., 2016) 017}, [\href{https://arxiv.org/abs/1505.07098}{{\ttfamily
  1505.07098}}].

\bibitem{Abidi:2018eyd}
M.~M. Abidi and T.~Baldauf, \emph{{Cubic Halo Bias in Eulerian and Lagrangian
  Space}}, \href{https://doi.org/10.1088/1475-7516/2018/07/029}{\emph{JCAP}
  {\bfseries 1807} (2018) 029},
  [\href{https://arxiv.org/abs/1802.07622}{{\ttfamily 1802.07622}}].

\bibitem{Lazeyras:2017hxw}
T.~Lazeyras and F.~Schmidt, \emph{{Beyond LIMD bias: a measurement of the
  complete set of third-order halo bias parameters}},
  \href{https://doi.org/10.1088/1475-7516/2018/09/008}{\emph{JCAP} {\bfseries
  1809} (2018) 008}, [\href{https://arxiv.org/abs/1712.07531}{{\ttfamily
  1712.07531}}].

\bibitem{2020arXiv200901200S}
T.~{Steele} and T.~{Baldauf}, \emph{{Precise Calibration of the One-Loop
  Bispectrum in the Effective Field Theory of Large Scale Structure}},
  {\emph{arXiv e-prints} (Sept., 2020) arXiv:2009.01200},
  [\href{https://arxiv.org/abs/2009.01200}{{\ttfamily 2009.01200}}].

\bibitem{gridSPT}
A.~Taruya, T.~Nishimichi and D.~Jeong, \emph{{Grid-based calculation for
  perturbation theory of large-scale structure}},
  \href{https://doi.org/10.1103/PhysRevD.98.103532}{\emph{Phys. Rev. D}
  {\bfseries 98} (2018) 103532},
  [\href{https://arxiv.org/abs/1807.04215}{{\ttfamily 1807.04215}}].

\bibitem{2020arXiv200705504T}
A.~{Taruya}, T.~{Nishimichi} and D.~{Jeong}, \emph{{The covariance of the
  matter power spectrum including the survey window function effect: N-body
  simulations vs. fifth-order perturbation theory on grid}}, {\emph{arXiv
  e-prints} (July, 2020) arXiv:2007.05504},
  [\href{https://arxiv.org/abs/2007.05504}{{\ttfamily 2007.05504}}].

\bibitem{2013MNRAS.432..894J}
J.~{Jasche} and B.~D. {Wandelt}, \emph{{Bayesian physical reconstruction of
  initial conditions from large-scale structure surveys}},
  \href{https://doi.org/10.1093/mnras/stt449}{\emph{\mnras} {\bfseries 432}
  (June, 2013) 894--913}, [\href{https://arxiv.org/abs/1203.3639}{{\ttfamily
  1203.3639}}].

\bibitem{2015MNRAS.446.4250A}
M.~{Ata}, F.-S. {Kitaura} and V.~{M{\"u}ller}, \emph{{Bayesian inference of
  cosmic density fields from non-linear, scale-dependent, and stochastic biased
  tracers}}, \href{https://doi.org/10.1093/mnras/stu2347}{\emph{\mnras}
  {\bfseries 446} (Feb., 2015) 4250--4259},
  [\href{https://arxiv.org/abs/1408.2566}{{\ttfamily 1408.2566}}].

\bibitem{2017JCAP...12..009S}
U.~{Seljak}, G.~{Aslanyan}, Y.~{Feng} and C.~{Modi}, \emph{{Towards optimal
  extraction of cosmological information from nonlinear data}},
  \href{https://doi.org/10.1088/1475-7516/2017/12/009}{\emph{\jcap} {\bfseries
  12} (Dec., 2017) 009}, [\href{https://arxiv.org/abs/1706.06645}{{\ttfamily
  1706.06645}}].

\bibitem{paperII}
F.~Elsner, F.~Schmidt, J.~Jasche, G.~Lavaux and N.-M. Nguyen, \emph{{Cosmology
  Inference from Biased Tracers using the EFT-based Likelihood}},
  \href{https://doi.org/10.1088/1475-7516/2020/01/029}{\emph{JCAP} {\bfseries
  2001} (2020) 029}, [\href{https://arxiv.org/abs/1906.07143}{{\ttfamily
  1906.07143}}].

\bibitem{2019arXiv190906396L}
G.~{Lavaux}, J.~{Jasche} and F.~{Leclercq}, \emph{{Systematic-free inference of
  the cosmic matter density field from SDSS3-BOSS data}}, {\emph{arXiv
  e-prints} (Sept., 2019) arXiv:1909.06396},
  [\href{https://arxiv.org/abs/1909.06396}{{\ttfamily 1909.06396}}].

\bibitem{LSSreview}
F.~Bernardeau, S.~Colombi, E.~Gaztanaga and R.~Scoccimarro, \emph{{Large scale
  structure of the universe and cosmological perturbation theory}},
  \href{https://doi.org/10.1016/S0370-1573(02)00135-7}{\emph{Phys.Rept.}
  {\bfseries 367} (2002) 1--248},
  [\href{https://arxiv.org/abs/astro-ph/0112551}{{\ttfamily
  astro-ph/0112551}}].

\bibitem{biasreview}
V.~{Desjacques}, D.~{Jeong} and F.~{Schmidt}, \emph{{Large-scale galaxy bias}},
  \href{https://doi.org/10.1016/j.physrep.2017.12.002}{\emph{\physrep}
  {\bfseries 733} (Feb., 2018) 1--193},
  [\href{https://arxiv.org/abs/1611.09787}{{\ttfamily 1611.09787}}].

\bibitem{schmittfull/etal:2018}
M.~Schmittfull, M.~Simonović, V.~Assassi and M.~Zaldarriaga, \emph{{Modeling
  Biased Tracers at the Field Level}},
  \href{https://doi.org/10.1103/PhysRevD.100.043514}{\emph{Phys. Rev.}
  {\bfseries D100} (2019) 043514},
  [\href{https://arxiv.org/abs/1811.10640}{{\ttfamily 1811.10640}}].

\bibitem{schmittfull/etal:2020}
M.~{Schmittfull}, M.~{Simonovi{\'c}}, M.~M. {Ivanov}, O.~H.~E. {Philcox} and
  M.~{Zaldarriaga}, \emph{{Modeling Galaxies in Redshift Space at the Field
  Level}}, {\emph{arXiv e-prints} (Dec., 2020) arXiv:2012.03334},
  [\href{https://arxiv.org/abs/2012.03334}{{\ttfamily 2012.03334}}].

\bibitem{fujita/vlah}
T.~{Fujita} and Z.~{Vlah}, \emph{{Perturbative description of biased tracers
  using consistency relations of LSS}},
  \href{https://doi.org/10.1088/1475-7516/2020/10/059}{\emph{\jcap} {\bfseries
  2020} (Oct., 2020) 059}, [\href{https://arxiv.org/abs/2003.10114}{{\ttfamily
  2003.10114}}].

\bibitem{2020JCAP...10..039D}
Y.~{Donath} and L.~{Senatore}, \emph{{Biased tracers in redshift space in the
  EFTofLSS with exact time dependence}},
  \href{https://doi.org/10.1088/1475-7516/2020/10/039}{\emph{\jcap} {\bfseries
  2020} (Oct., 2020) 039}, [\href{https://arxiv.org/abs/2005.04805}{{\ttfamily
  2005.04805}}].

\bibitem{1992MNRAS.254..729B}
T.~{Buchert}, \emph{{Lagrangian theory of gravitational instability of
  Friedman-Lemaitre cosmologies and the 'Zel'dovich approximation'}},
  \href{https://doi.org/10.1093/mnras/254.4.729}{\emph{\mnras} {\bfseries 254}
  (Feb., 1992) 729--737}.

\bibitem{1994MNRAS.267..811B}
T.~{Buchert}, \emph{{Lagrangian Theory of Gravitational Instability of
  Friedman-Lemaitre Cosmologies - a Generic Third-Order Model for Nonlinear
  Clustering}}, \href{https://doi.org/10.1093/mnras/267.4.811}{\emph{\mnras}
  {\bfseries 267} (Apr., 1994) 811},
  [\href{https://arxiv.org/abs/astro-ph/9309055}{{\ttfamily
  astro-ph/9309055}}].

\bibitem{1995A&A...296..575B}
F.~R. {Bouchet}, S.~{Colombi}, E.~{Hivon} and R.~{Juszkiewicz},
  \emph{{Perturbative Lagrangian approach to gravitational instability.}},
  {\emph{\aap} {\bfseries 296} (Apr., 1995) 575},
  [\href{https://arxiv.org/abs/astro-ph/9406013}{{\ttfamily
  astro-ph/9406013}}].

\bibitem{1995MNRAS.276..115C}
P.~{Catelan}, \emph{{Lagrangian dynamics in non-flat universes and non-linear
  gravitational evolution}},
  \href{https://doi.org/10.1093/mnras/276.1.115}{\emph{\mnras} {\bfseries 276}
  (Sept., 1995) 115--124},
  [\href{https://arxiv.org/abs/astro-ph/9406016}{{\ttfamily
  astro-ph/9406016}}].

\bibitem{2012JCAP...06..021R}
C.~{Rampf} and T.~{Buchert}, \emph{{Lagrangian perturbations and the matter
  bispectrum I: fourth-order model for non-linear clustering}},
  \href{https://doi.org/10.1088/1475-7516/2012/06/021}{\emph{\jcap} {\bfseries
  2012} (June, 2012) 021}, [\href{https://arxiv.org/abs/1203.4260}{{\ttfamily
  1203.4260}}].

\bibitem{2015JCAP...09..014V}
Z.~{Vlah}, M.~{White} and A.~{Aviles}, \emph{{A Lagrangian effective field
  theory}}, \href{https://doi.org/10.1088/1475-7516/2015/09/014}{\emph{\jcap}
  {\bfseries 2015} (Sept., 2015) 014},
  [\href{https://arxiv.org/abs/1506.05264}{{\ttfamily 1506.05264}}].

\bibitem{matsubara:2008}
T.~{Matsubara}, \emph{{Nonlinear perturbation theory with halo bias and
  redshift-space distortions via the Lagrangian picture}},
  \href{https://doi.org/10.1103/PhysRevD.78.083519}{\emph{\prd} {\bfseries 78}
  (Oct., 2008) 083519}, [\href{https://arxiv.org/abs/0807.1733}{{\ttfamily
  0807.1733}}].

\bibitem{2020JCAP...07..062C}
S.-F. {Chen}, Z.~{Vlah} and M.~{White}, \emph{{Consistent modeling of velocity
  statistics and redshift-space distortions in one-loop perturbation theory}},
  \href{https://doi.org/10.1088/1475-7516/2020/07/062}{\emph{\jcap} {\bfseries
  2020} (July, 2020) 062}, [\href{https://arxiv.org/abs/2005.00523}{{\ttfamily
  2005.00523}}].

\bibitem{2014JCAP...06..008T}
S.~{Tassev}, \emph{{Lagrangian or Eulerian; real or Fourier? Not all approaches
  to large-scale structure are created equal}},
  \href{https://doi.org/10.1088/1475-7516/2014/06/008}{\emph{\jcap} {\bfseries
  2014} (June, 2014) 008}, [\href{https://arxiv.org/abs/1311.4884}{{\ttfamily
  1311.4884}}].

\bibitem{MSZ}
M.~{Mirbabayi}, F.~{Schmidt} and M.~{Zaldarriaga}, \emph{{Biased tracers and
  time evolution}},
  \href{https://doi.org/10.1088/1475-7516/2015/07/030}{\emph{\jcap} {\bfseries
  7} (July, 2015) 030}, [\href{https://arxiv.org/abs/1412.5169}{{\ttfamily
  1412.5169}}].

\bibitem{Cabass:2020jqo}
G.~Cabass, \emph{{The EFT Likelihood for Large-Scale Structure in Redshift
  Space}},  \href{https://arxiv.org/abs/2007.14988}{{\ttfamily 2007.14988}}.

\bibitem{cabass/schmidt:2020}
G.~Cabass and F.~Schmidt, \emph{{The Likelihood for LSS: Stochasticity of Bias
  Coefficients at All Orders}},
  \href{https://doi.org/10.1088/1475-7516/2020/07/051}{\emph{JCAP} {\bfseries
  07} (2020) 051}, [\href{https://arxiv.org/abs/2004.00617}{{\ttfamily
  2004.00617}}].

\bibitem{paperI}
F.~{Schmidt}, F.~{Elsner}, J.~{Jasche}, N.~M. {Nguyen} and G.~{Lavaux},
  \emph{{A rigorous EFT-based forward model for large-scale structure}},
  \href{https://doi.org/10.1088/1475-7516/2019/01/042}{\emph{{Journal of
  Cosmology and Astro-Particle Physics}} {\bfseries 2019} (Jan, 2019) 042},
  [\href{https://arxiv.org/abs/1808.02002}{{\ttfamily 1808.02002}}].

\bibitem{paperIIb}
F.~{Schmidt}, G.~{Cabass}, J.~{Jasche} and G.~{Lavaux}, \emph{{Unbiased
  cosmology inference from biased tracers using the EFT likelihood}},
  \href{https://doi.org/10.1088/1475-7516/2020/11/008}{\emph{\jcap} {\bfseries
  2020} (Nov., 2020) 008}, [\href{https://arxiv.org/abs/2004.06707}{{\ttfamily
  2004.06707}}].

\bibitem{cabass/schmidt:2019}
G.~Cabass and F.~Schmidt, \emph{{The EFT Likelihood for Large-Scale
  Structure}}, \href{https://doi.org/10.1088/1475-7516/2020/04/042}{\emph{JCAP}
  {\bfseries 04} (2020) 042},
  [\href{https://arxiv.org/abs/1909.04022}{{\ttfamily 1909.04022}}].

\bibitem{jeong/schmidt:2015}
D.~{Jeong} and F.~{Schmidt}, \emph{{Large-scale structure observables in
  general relativity}},
  \href{https://doi.org/10.1088/0264-9381/32/4/044001}{\emph{Classical and
  Quantum Gravity} {\bfseries 32} (Feb., 2015) 044001},
  [\href{https://arxiv.org/abs/1407.7979}{{\ttfamily 1407.7979}}].

\bibitem{rampf:2012}
C.~{Rampf}, \emph{{The recursion relation in Lagrangian perturbation theory}},
  \href{https://doi.org/10.1088/1475-7516/2012/12/004}{\emph{\jcap} {\bfseries
  12} (Dec., 2012) 4}, [\href{https://arxiv.org/abs/1205.5274}{{\ttfamily
  1205.5274}}].

\bibitem{zheligovsky/frisch}
V.~{Zheligovsky} and U.~{Frisch}, \emph{{Time-analyticity of Lagrangian
  particle trajectories in ideal fluid flow}},
  \href{https://doi.org/10.1017/jfm.2014.221}{\emph{Journal of Fluid Mechanics}
  {\bfseries 749} (June, 2014) 404--430},
  [\href{https://arxiv.org/abs/1312.6320}{{\ttfamily 1312.6320}}].

\bibitem{matsubara:2015}
T.~{Matsubara}, \emph{{Recursive solutions of Lagrangian perturbation theory}},
  \href{https://doi.org/10.1103/PhysRevD.92.023534}{\emph{\prd} {\bfseries 92}
  (July, 2015) 023534}, [\href{https://arxiv.org/abs/1505.01481}{{\ttfamily
  1505.01481}}].

\bibitem{buchert/ehlers:1997}
T.~{Buchert} and J.~{Ehlers}, \emph{{Averaging inhomogeneous Newtonian
  cosmologies.}}, {\emph{\aap} {\bfseries 320} (Apr., 1997) 1--7},
  [\href{https://arxiv.org/abs/astro-ph/9510056}{{\ttfamily
  astro-ph/9510056}}].

\bibitem{1997GReGr..29..733E}
J.~{Ehlers} and T.~{Buchert}, \emph{{Newtonian Cosmology in Lagrangian
  Formulation: Foundations and Perturbation Theory}},
  \href{https://doi.org/10.1023/A:1018885922682}{\emph{General Relativity and
  Gravitation} {\bfseries 29} (June, 1997) 733--764},
  [\href{https://arxiv.org/abs/astro-ph/9609036}{{\ttfamily
  astro-ph/9609036}}].

\bibitem{crocce/etal:2006}
M.~{Crocce}, S.~{Pueblas} and R.~{Scoccimarro}, \emph{{Transients from initial
  conditions in cosmological simulations}},
  \href{https://doi.org/10.1111/j.1365-2966.2006.11040.x}{\emph{\mnras}
  {\bfseries 373} (Nov., 2006) 369--381},
  [\href{https://arxiv.org/abs/astro-ph/0606505}{{\ttfamily
  astro-ph/0606505}}].

\bibitem{MUSIC}
O.~{Hahn}, M.~{Michaux}, C.~{Rampf}, C.~{Uhlemann} and R.~E. {Angulo},
  \emph{{MUSIC2-monofonIC: 3LPT initial condition generator}},  Aug., 2020.

\bibitem{2020arXiv201012584R}
C.~{Rampf} and O.~{Hahn}, \emph{{Shell-crossing in a $\Lambda$CDM Universe}},
  {\emph{arXiv e-prints} (Oct., 2020) arXiv:2010.12584},
  [\href{https://arxiv.org/abs/2010.12584}{{\ttfamily 2010.12584}}].

\bibitem{senatore:2015}
L.~{Senatore}, \emph{{Bias in the effective field theory of large scale
  structures}},
  \href{https://doi.org/10.1088/1475-7516/2015/11/007}{\emph{\jcap} {\bfseries
  11} (Nov., 2015) 007}, [\href{https://arxiv.org/abs/1406.7843}{{\ttfamily
  1406.7843}}].

\bibitem{fujita/etal}
T.~{Fujita}, V.~{Mauerhofer}, L.~{Senatore}, Z.~{Vlah} and R.~{Angulo},
  \emph{{Very massive tracers and higher derivative biases}},
  \href{https://doi.org/10.1088/1475-7516/2020/01/009}{\emph{\jcap} {\bfseries
  2020} (Jan., 2020) 009}, [\href{https://arxiv.org/abs/1609.00717}{{\ttfamily
  1609.00717}}].

\bibitem{lazeyras/schmidt}
T.~{Lazeyras} and F.~{Schmidt}, \emph{{A robust measurement of the first
  higher-derivative bias of dark matter halos}}, {\emph{arXiv e-prints} (Apr,
  2019) arXiv:1904.11294}, [\href{https://arxiv.org/abs/1904.11294}{{\ttfamily
  1904.11294}}].

\bibitem{2007JCAP...10..007C}
P.~{Coles} and P.~{Erdogdu}, \emph{{Scale dependent galaxy bias}},
  \href{https://doi.org/10.1088/1475-7516/2007/10/007}{\emph{\jcap} {\bfseries
  2007} (Oct., 2007) 007}, [\href{https://arxiv.org/abs/0706.0412}{{\ttfamily
  0706.0412}}].

\bibitem{2014PhRvD..89h3010P}
A.~{Pontzen}, \emph{{Scale-dependent bias in the
  baryonic-acoustic-oscillation-scale intergalactic neutral hydrogen}},
  \href{https://doi.org/10.1103/PhysRevD.89.083010}{\emph{\prd} {\bfseries 89}
  (Apr., 2014) 083010}, [\href{https://arxiv.org/abs/1402.0506}{{\ttfamily
  1402.0506}}].

\bibitem{2019JCAP...05..031C}
G.~{Cabass} and F.~{Schmidt}, \emph{{A new scale in the bias expansion}},
  \href{https://doi.org/10.1088/1475-7516/2019/05/031}{\emph{\jcap} {\bfseries
  2019} (May, 2019) 031}, [\href{https://arxiv.org/abs/1812.02731}{{\ttfamily
  1812.02731}}].

\bibitem{Orszag}
S.~A. {Orszag}, \emph{{On the Elimination of Aliasing in Finite-Difference
  Schemes by Filtering High-Wavenumber Components.}},
  \href{https://doi.org/10.1175/1520-0469(1971)028<1074:OTEOAI>2.0.CO;2}{\emph{Journal
  of Atmospheric Sciences} {\bfseries 28} (Sept., 1971) 1074--1074}.

\bibitem{2005MNRAS.364.1105S}
V.~{Springel}, \emph{{The cosmological simulation code GADGET-2}},
  \href{https://doi.org/10.1111/j.1365-2966.2005.09655.x}{\emph{\mnras}
  {\bfseries 364} (Dec., 2005) 1105--1134},
  [\href{https://arxiv.org/abs/astro-ph/0505010}{{\ttfamily
  astro-ph/0505010}}].

\bibitem{coles/etal:1993}
P.~{Coles}, A.~L. {Melott} and S.~F. {Shandarin}, \emph{{Testing approximations
  for non-linear gravitational clustering}},
  \href{https://doi.org/10.1093/mnras/260.4.765}{\emph{\mnras} {\bfseries 260}
  (Feb., 1993) 765--776}.

\bibitem{rampf/villone/frisch}
C.~{Rampf}, B.~{Villone} and U.~{Frisch}, \emph{{How smooth are particle
  trajectories in a {\ensuremath{\Lambda}}CDM Universe?}},
  \href{https://doi.org/10.1093/mnras/stv1365}{\emph{\mnras} {\bfseries 452}
  (Sept., 2015) 1421--1436},
  [\href{https://arxiv.org/abs/1504.00032}{{\ttfamily 1504.00032}}].

\bibitem{mccullagh/etal}
N.~{McCullagh}, D.~{Jeong} and A.~S. {Szalay}, \emph{{Toward accurate modelling
  of the non-linear matter bispectrum: standard perturbation theory and
  transients from initial conditions}},
  \href{https://doi.org/10.1093/mnras/stv2525}{\emph{\mnras} {\bfseries 455}
  (Jan., 2016) 2945--2958}, [\href{https://arxiv.org/abs/1507.07824}{{\ttfamily
  1507.07824}}].

\bibitem{Nishimichi:2018etk}
T.~Nishimichi et~al., \emph{{Dark Quest. I. Fast and Accurate Emulation of Halo
  Clustering Statistics and Its Application to Galaxy Clustering}},
  \href{https://doi.org/10.3847/1538-4357/ab3719}{\emph{Astrophys. J.}
  {\bfseries 884} (2019) 29},
  [\href{https://arxiv.org/abs/1811.09504}{{\ttfamily 1811.09504}}].

\bibitem{Michaux:2020yis}
M.~Michaux, O.~Hahn, C.~Rampf and R.~E. Angulo, \emph{{Accurate initial
  conditions for cosmological N-body simulations: Minimizing truncation and
  discreteness errors}},  \href{https://arxiv.org/abs/2008.09588}{{\ttfamily
  2008.09588}}.

\bibitem{baumann/etal:2012}
D.~{Baumann}, A.~{Nicolis}, L.~{Senatore} and M.~{Zaldarriaga},
  \emph{{Cosmological non-linearities as an effective fluid}},
  \href{https://doi.org/10.1088/1475-7516/2012/07/051}{\emph{\jcap} {\bfseries
  7} (July, 2012) 051}, [\href{https://arxiv.org/abs/1004.2488}{{\ttfamily
  1004.2488}}].

\bibitem{carrasco/etal:2012}
J.~J.~M. {Carrasco}, M.~P. {Hertzberg} and L.~{Senatore}, \emph{{The effective
  field theory of cosmological large scale structures}},
  \href{https://doi.org/10.1007/JHEP09(2012)082}{\emph{Journal of High Energy
  Physics} {\bfseries 9} (Sept., 2012) 82},
  [\href{https://arxiv.org/abs/1206.2926}{{\ttfamily 1206.2926}}].

\bibitem{carroll/etal}
S.~M. Carroll, S.~Leichenauer and J.~Pollack, \emph{{Consistent effective
  theory of long-wavelength cosmological perturbations}},
  \href{https://doi.org/10.1103/PhysRevD.90.023518}{\emph{Phys. Rev.}
  {\bfseries D90} (2014) 023518},
  [\href{https://arxiv.org/abs/1310.2920}{{\ttfamily 1310.2920}}].

\bibitem{Carrasco:2013sva}
J.~J.~M. Carrasco, S.~Foreman, D.~Green and L.~Senatore, \emph{{The 2-loop
  matter power spectrum and the IR-safe integrand}},
  \href{https://doi.org/10.1088/1475-7516/2014/07/056}{\emph{JCAP} {\bfseries
  1407} (2014) 056}, [\href{https://arxiv.org/abs/1304.4946}{{\ttfamily
  1304.4946}}].

\bibitem{Angulo:2015}
R.~Angulo, M.~Fasiello, L.~Senatore and Z.~Vlah, \emph{{On the Statistics of
  Biased Tracers in the Effective Field Theory of Large Scale Structures}},
  \href{https://doi.org/10.1088/1475-7516/2015/09/029,
  10.1088/1475-7516/2015/9/029}{\emph{JCAP} {\bfseries 1509} (2015) 029},
  [\href{https://arxiv.org/abs/1503.08826}{{\ttfamily 1503.08826}}].

\bibitem{baldauf/etal:2015}
T.~Baldauf, L.~Mercolli and M.~Zaldarriaga, \emph{{Effective field theory of
  large scale structure at two loops: The apparent scale dependence of the
  speed of sound}},
  \href{https://doi.org/10.1103/PhysRevD.92.123007}{\emph{Phys. Rev.}
  {\bfseries D92} (2015) 123007},
  [\href{https://arxiv.org/abs/1507.02256}{{\ttfamily 1507.02256}}].

\bibitem{buchert/etal:1994}
T.~{Buchert}, A.~L. {Melott} and A.~G. {Weiss}, \emph{{Testing higher-order
  Lagrangian perturbation theory against numerical simulations I. Pancake
  models}}, {\emph{\aap} {\bfseries 288} (Aug., 1994) 349--364},
  [\href{https://arxiv.org/abs/astro-ph/9309056}{{\ttfamily
  astro-ph/9309056}}].

\bibitem{melott/etal:1995}
A.~L. {Melott}, T.~{Buchert} and A.~G. {Weiss}, \emph{{Testing higher-order
  Lagrangian perturbation theory against numerical simulations. II.
  Hierarchical models.}}, {\emph{\aap} {\bfseries 294} (Feb., 1995) 345--365},
  [\href{https://arxiv.org/abs/astro-ph/9404018}{{\ttfamily
  astro-ph/9404018}}].

\bibitem{2017arXiv170901544S}
F.~{Schmidt}, \emph{{Monodromic Dark Energy}}, {\emph{arXiv e-prints} (Sept.,
  2017) arXiv:1709.01544}, [\href{https://arxiv.org/abs/1709.01544}{{\ttfamily
  1709.01544}}].

\bibitem{2018arXiv180807496K}
D.~K. {Ramanah}, G.~{Lavaux}, J.~{Jasche} and B.~D. {Wandelt},
  \emph{{Cosmological inference from Bayesian forward modelling of deep galaxy
  redshift surveys}},
  \href{https://doi.org/10.1051/0004-6361/201834117}{\emph{\aap} {\bfseries
  621} (Jan, 2019) A69}, [\href{https://arxiv.org/abs/1808.07496}{{\ttfamily
  1808.07496}}].

\end{thebibliography}\endgroup

\end{document}